\documentclass[lettersize,journal]{IEEEtran}
\usepackage{amsmath,amsfonts}
\usepackage{algorithmic}
\usepackage{algorithm}
\usepackage{array}
\usepackage[caption=false,font=normalsize,labelfont=sf,textfont=sf]{subfig}
\usepackage{textcomp}
\usepackage{stfloats}
\usepackage{url}
\usepackage{verbatim}
\usepackage{graphicx}
\usepackage{cite}
\usepackage{booktabs}
\usepackage{tikz}
\usepackage{import}
\usepackage{flushend}  %评审最后一页

\begin{document}

\title{Review of Passenger Flow Modelling Approaches\\Based on a Bibliometric Analysis}

\author{Jonathan Hecht, Weilian Li, Ziyue Li, Youness Dehbi
        % <-this % stops a space
\thanks{This paper was supported by the Federal Ministry of Transport and Digital Infrastructure (BMVI) under Grant 19OI22008A. \textit{(Corresponding author: Weilian Li)}}% <-this % stops a space
\thanks{Jonathan Hecht, Weilian Li, and Youness Dehbi are with the Computa-
tional Methods Lab, HafenCity University Hamburg, Hamburg 20457, Germany}
\thanks{Weilian Li is with the Faculty of Geosciences and Engineering, Southwest Jiaotong University, Chengdu 611756, China}
\thanks{Ziyue Li is with the Heilbronn Data Science Center and Department of Operations \&Technology, Technical University of Munich, 80333 Munich, Germany}
% \thanks{Manuscript received April 19, 2021; revised August 16, 2021.}}

}

% The paper headers
\markboth{IEEE TRANSACTIONS ON INTELLIGENT TRANSPORTATION SYSTEMS, VOL. XX, NO. XX, DECEMBER 202X}%
{Shell \MakeLowercase{\textit{et al.}}: A Sample Article Using IEEEtran.cls for IEEE Journals}

% Remember, if you use this you must call \IEEEpubidadjcol in the second
% column for its text to clear the IEEEpubid mark.

\maketitle

\begin{abstract}
This paper presents a bibliometric analysis of the field of short-term passenger flow forecasting within local public transit, covering 814 publications that span from 1984 to 2024. In addition to common bibliometric analysis tools, a variant of a citation network was developed, and topic modelling was conducted. The analysis reveals that research activity exhibited sporadic patterns prior to 2008, followed by a marked acceleration, characterised by a shift from conventional statistical and machine learning methodologies (e.g., ARIMA, SVM, and basic neural networks) to specialised deep learning architectures. Based on this insight, a connection to more general fields such as machine learning and time series modelling was established. In addition to modelling, spatial, linguistic, and modal biases were identified and findings from existing secondary literature were validated and quantified. This revealed existing gaps, such as constrained data fusion, open (multivariate) data, and underappreciated challenges related to model interpretability, cost-efficiency, and a balance between algorithmic performance and practical deployment considerations. In connection with the superordinate fields, the growth in relevance of foundation models is also noteworthy.
\end{abstract}


\section{Introduction}
\label{sec:introduction}
\IEEEPARstart{P}{ublic} transit systems constitute a fundamental component of urban infrastructure, offering substantial capacities, punctuality, speed, and environmental advantages \cite{Kwan.2016, Thilakaratne.2011, Vuchic.2007}. Consequently, passenger flow forecasting emerges as an essential instrument to address the requirements and expectations of public transit users, contributing significantly to the strategic planning and operation of intelligent transportation systems \cite{Dimitrakopoulos.2010}. Specifically, accurate data on the number of passengers present in the system at any given moment and in the near future, commonly referred to as short-term passenger flow prediction, are crucial for informed decision-making. This information enables dispatchers, for example, to implement suitable countermeasures in response to cancellations affecting current and anticipated passengers. Furthermore, security personnel could pre-emptively restrict access to tracks or stations to mitigate the risks associated with overcrowding. In addition, passengers might be advised on alternative or later connections. Various modelling methodologies have been used in the domain of passenger flow forecasting. This paper presents a bibliometric analysis of the existing scholarly literature with an emphasis on short-term passenger flow modelling for public transit, building on and further expanding the existing secondary literature.

Prior to outlining existing review papers, the understanding of passenger flow in local public transit should be briefly re-examined. Initially, a distinction is drawn between the prediction intervals. The term 'short-term' typically encompasses intervals of less than an hour, whereas 'long-term' refers to forecasts of annual or monthly passenger flow, which are utilised for planning and designing purposes \cite{Xue.2023}. In the area of long-term passenger flow and more general traffic forecasting, for example, the four-step model can be seen \cite{McNALLY.2000}. Another distinction pertains to the various levels of forecast. Passenger flow predictions can be made for the entire network, line, or station level within a transit system \cite{Xue.2023}. Furthermore, when focusing on the objectives of the modelling, at least four categories of passenger flows can be identified. Specifically, inbound, outbound, and transfer flows can be differentiated at the station level, while origin-destination passenger flows are considered primarily at the network level \cite{Xue.2023}. Lastly, \cite{Xue.2023} articulated four modelling scenarios: These encompass the aforementioned short-term passenger flow forecasting and long-term passenger flow forecasting, in addition to passenger flow forecasts during emergencies and those for a typical day. In this study, public transit is defined, following \cite{Ceder.2020}, as any travel system available for public use, including both conventional vehicles and autonomous vehicles. This broad definition also encompasses the terms public transport, public transportation, mass transit, or simply transit.

The aforementioned scenarios contribute to the complexity exceeding that encountered in other time series forecasting challenges. In addition to the different dynamics of the scenarios, the amount of data, the high dimensionality, and the spatio-temporal relationship of the data sets are challenging \cite{Jiang.2022}. In spite of these complexities, it is feasible to formally define our objective. This objective is to forecast future passenger flows using historical data. Given a univariate time series $\{x_t\}$, where $x_t$ denotes the observed passenger flow at time $t$, the objective is to estimate future values $\{x_{t+1}, x_{t+2}, \dots, x_{t+H}\}$ over a given prediction horizon $H$ based on previous observations $\{x_{t-n}, \dots, x_{t-1}, x_t\}$. Formally, this problem can be expressed as: 
\begin{align*}
    &\hat{x}_{t+h} = f(x_{t-n}, x_{t-n+1}, \dots, x_t), \quad \\
    &\text{for } h = 1, 2, \dots, H,
\end{align*}
where: \begin{itemize}
    \item $\hat{x}_{t+h}$ is the predicted passenger flow at time $t+h$.
    \item $f$ is the predictive model.
    \item $n$ is the size of the input sequence, representing the number of previous observations used for the prediction.
    \item $H$ is the forecasting horizon.
\end{itemize}

The notion of modelling passenger flows as a time series is equally applicable to other disciplines, such as the general field of traffic forecasting. Consequently, the ensuing literature review considers elements of existing analyses on general traffic modelling in a summarised form, followed by a comprehensive examination of reviews related to local public transit.\\
In terms of traffic forecasts, \cite{Vlahogianni.2004} conducted an analysis of the entire range of possibilities to predict various traffic parameters, as documented in the literature up to 2003. The analysis proposed a basic framework for the development of traffic forecasting algorithms based on conceptual properties. In addition to identifying model types also employed in the domain of local public transit, the differentiation into parametric and nonparametric models is noteworthy. Similarly, \cite{Vlahogianni.2014} identified ten challenges in the literature over the following decade, several of which remain relevant and can be adapted to the forecasting of passenger flow. For instance, challenge five, "Using new technologies for collecting and fusing data," highlighted, among other considerations, the interplay between privacy issues and data integration.
\begin{table*}[b]
    \caption{Overview of Selected Studies on Passenger Flow Forecasting.}
    \label{tab:passenger_flow}
    \normalsize
    \centering
    \renewcommand{\arraystretch}{1.5}
    \resizebox{\textwidth}{!}{
    \begin{tabular}{p{4cm} p{4cm} p{1.5cm} p{4cm} p{4cm} p{1.5cm} p{4cm} p{4cm} p{1.5cm}}
        \hline
        Author(s) & Title & Publication Year & Methodological Approach & Research Object / Focus & \# Included Papers & Main Selection Criteria & Data Sources & Temporal Coverage \\
        \hline
        Eleni I. Vlahogianni, Matthew G. Karlaftis, John Golias & Short-term traffic forecasting: Where we are and where we’re going & 2014 & Not explicitly mentioned (Qualitative approach)  & Discussion about challenges in short-term traffic forecasting & 128 & — & — & 2004-2013\\
        Weiwei Jiang, Jiayun Luo & Graph Neural Network for Traffic Forecasting: A survey & 2022 & ot explicitly mentioned (Qualitative approach)  & Review of graph neural networks for various traffic forecasting problems& 212& — & — & 2018-2020\\
        Huawei Zhai, Licheng Cui, Yu Nie, Xiaowei Xu, Weishi Zhang & A Comprehensive Comparative Analysis of the Basic Theory of the Short Term Bus Passenger Flow Prediction & 2018 & Not explicitly mentioned (Qualitative approach) & Short-term bus passenger flow forecasting models & 20 & English + Chinese; Keywords: short term + bus passenger + ridership + predict + forecast; Not Urban Rail Transit & Among others: Science Citation Index, Engineering Index, ScienceDirect, IEEE Xplore Digital Library, China Knowledge Resource Integrated Database & 2000-2017 \\
        Etikaf Hussain, Ashish Bhaskar, Edward Chung & Transit OD matrix estimation using smartcard data: Recent developments and future research challenges & 2021 & Not explicitly mentioned (Qualitative approach) & Literature review on transit Origin-Destination estimation using smartcard data & 31 & — & — & 2002-2020 \\
        Qiuchi Xue, Wei Zhang, Meiling Ding, Xin Yang, Jianjun Wu, Ziyou Gao & Passenger flow forecasting approaches for urban rail transit: a survey & 2023 & Not explicitly mentioned (Qualitative approach) & Review of forecasting methods for passenger flow on URT within four different scenarios & ($\approx$58) & Inbound or outbound flow/volume forecasting & — & ($\approx$2014-2023) \\
        Franca Rocco di Torrepadula, Enea Vincenzo Napolitano, Sergio Di Martino, Nicola Mazzocca & Machine Learning for public transportation demand prediction: A Systematic Literature Review & 2024 & Systematic Literature Review + Systematic Mapping Study & ML techniques for predicting public transportation demand & 253 & Excluded: fundamental spatio-temporal predictive techniques, traffic-flow predictions; English; Field: engineering or Computer Science & Scopus, IEEE Xplore, ACM Digital Library & 2015-2023 \\
        Our Names & Review of Passenger Flow Forecast Approaches Based on a Bibliometric Analysis & 2025 & Bibliometric Analysis & Bibliometric analysis of approaches to model public transit passenger flows & 814 & English; All public transit modes & Web of Science, Scopus & 1984-2024 \\
        \hline
    \end{tabular}
    }
\end{table*}

More recently, \cite{Jiang.2022} conducted a survey on graph neural networks for traffic forecasting, incorporating various forms of passenger flow forecasting. Their survey is more comprehensive, covering all modalities and different prediction parameters such as travel time or speed, yet it is more focused on the exploration of prediction methods.

In general, four publications have been identified that solely examine passenger flows within public transit from different points of view. The first publication by \cite{Zhai.2018} focuses on the short-term flow of bus passengers, analysing the methodology of 20 selected journal articles sourced from scientific databases. For comparative purposes, it distinguishes among three categories of models: linear models (e.g., ARIMA), nonlinear models (e.g,. SVM), and combined models, thus expanding the categories by \cite{Vlahogianni.2004} and changing from (non)parametric to (non)linear, which reveals a different understanding of the models. The article focuses exclusively on the bus mode, which is responsible for the limited number of publications reviewed. Furthermore, considerable advancements have occurred in the field of modelling up to the present date. In contrast, \cite{Hussain.2021} explored the potential for estimating transit origin-destination matrices. The study focuses on the creation process and uses literature to investigate possible sources of error. Predictions of passenger transit flow are only one possible application. Thus, modelling these processes constitutes only a minor component of the examination.

In contrast to the previously mentioned research, \cite{Xue.2023} and \cite{Rocco.2024} have stronger overlaps with this bibliometric analysis. Consequently, our findings are later integrated and synthesised with their results. \cite{Xue.2023} initially delineates the framework of passenger flow tasks, encompassing the forecast content, available data, the forecast scenario, the forecast approach and the forecast indicators, with the corresponding subcategories elaborated in detail. An illustration of this is the categorisation of different flow types within the forecast content. A qualitative review is undertaken for each of these subcategories, although the method of literature selection and identification remains undisclosed. The most recent study by \cite{Rocco.2024} performs a systematic review of the literature on public transportation demand, using machine learning (ML) techniques during the period 2015-2023. In their study, the focus was only on ML, excluding other approaches. A comparison of the different publications can be found in Table~\ref{tab:passenger_flow}.

In summary, existing studies are predominantly based on qualitative analyses, encompass shorter temporal scopes, or focus exclusively on a specific mode. Moreover, disparate target variables are occasionally amalgamated within a single study. Furthermore, there are limited connections made to emerging trends and, consequently, to prospective future modelling methodologies. Additionally, contemporary research lacks the integration of other reviews. To address the shortcomings in extant reviews concerning passenger flow in local public transit, the ensuing questions are examined:
\begin{enumerate} 
    \item How have the topics, focus points, and general structure (including researchers, dominant modes, and modelling approaches) evolved in the field of local public transit passenger flow forecasting? 
    \item How do these aspects inform potential future methodologies within a broader context? 
    \item To what extent can existing insights and approaches be substantiated, and how do they align with the anticipated future advancements? 
\end{enumerate}

\begin{figure}[thb]
    \centering
    \includegraphics[width=\columnwidth]{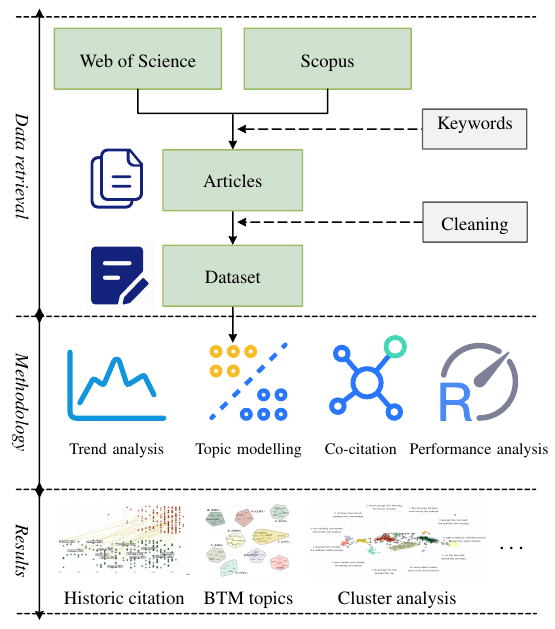}
    \caption{Illustration of the approach used in this study. The steps taken begin with data collection in combination with pre-processing, followed by various forms of analysis techniques, which lead to results that allow conclusions to be drawn about passenger flow modelling.}
    \label{fig:workflow}
\end{figure}

An overview of the approach used to address the questions is provided in Fig. \ref{fig:workflow}. Consequently, this paper seeks to offer a quantitative overview of passenger flow prediction modelling approaches, incorporating pre-existing findings. The choice of different modes of local public transport is not limited, just as no explicit search for long-term or long-distance prediction periods is performed. The paper is structured as follows: initially, it delineates the undertaking of the bibliometric analysis and subsequent synthesis. This is succeeded by presenting the results of the bibliometric analysis and their integration into existing reviews. Finally, the findings are contextualised within the discussion, and a conclusion is formulated.

%\begin{figure}[thb]
%\centering
%\resizebox{\columnwidth}{!}{\import{fig/}{method.pdf_tex}}
%  \caption{Illustration of the approach used in this study. The steps taken begin with data collection in combination with pre-processing, followed by various forms of analysis techniques, which lead to results that allow conclusions to be drawn about passenger flow modelling.}
%  \label{fig:workflow}
%\end{figure}

\section{Methodology}
\noindent Against the background of the research questions, various methodological approaches were initially considered, including different forms of systematic reviews or meta-analysis \cite{Sutton.2019, Pigott.2020}. Given the research questions and experimental queries, most approaches were deemed inapplicable due to the scope of the review and the number of documents that appear inexpedient for a complete manual review \cite{Donthu.2021}. To adequately address the questions raised, a bibliometric analysis is combined with a qualitative synthesis of existing review articles.  Bibliometric analysis enables the identification of knowledge gaps, the identification of new research approaches, and the obtainment of a broad overview of the entire field \cite{Donthu.2021}. The subsequent stage allows for the incorporation of one's findings with existing research, thereby generating additional insights.

To quantitatively assess the dominant themes and advancements in the domain of short-term passenger flow modelling in public transit, the bibliometric analysis procedure outlined in \cite{Donthu.2021} is adhered to as a guideline. These guidelines suggest first defining the objectives and scope, followed by the selection of suitable analytical techniques. This is followed by the definition of search terms and databases to create the data foundation. Finally, details on the implementation are presented.

\subsection{Objectives and Scope}
\noindent The scope and objectives have already been addressed in the Introduction. Specifically, the bibliographic analysis seeks to encompass the full spectrum of techniques to model passenger flows in local public transit systems. It is not constrained by any single period of publication, nor does it explicitly exclude any modelling time span. However, as will become evident later, no explicit exploration is conducted for long-term predictions, which could potentially be identified using alternative search terms. Thus, emphasis is placed on short-term forecasting techniques. Furthermore, the question aims more at an overview of content, including temporal trends, than at the interconnections between research institutions. 

\subsection{Analysis Techniques}
\noindent Among the various methodologies used in bibliometric analysis, a distinction is frequently drawn between the subareas of performance analysis and science mapping \cite{Kumar.2023, Passas.2024}. Generally, performance analysis refers to the evaluation of contributions from various research constituents, including authors, countries, or journals. In contrast, science mapping encompasses the examination of the interconnections between research constituents \cite{Donthu.2021, Kumar.2023}.

Within this framework, performance analysis primarily involves metrics associated with publications to assess the performance of various research actors, such as the number of publications by country, the duration of active research years or the publication channels. After an initial descriptive overview of the field has been obtained, science mapping is used to concretise the relationships between the research actors, in particular, content-related topics and developments. The techniques thus employed include the analysis of co-citations in various forms, as well as diverse co-word analysis methodologies. The usage of a biterm topic model for a deeper understanding of the relationships can be emphasised in the co-word analysis \cite{Yan.2013}.

Beyond methods applicable to bibliometric analysis, a topic model was developed from the document abstracts to furnish a more exhaustive overview. The specific model implemented was BERTopic, which utilizes pre-trained transformer-based language models to produce document embeddings. These embeddings are subsequently clustered and summarised into coherent topic representations through a class-based variant of Term Frequency–Inverse Document Frequency \cite{Grootendorst.2022}.

\subsection{Data Collection}
\noindent Following the selection of the appropriate techniques, the relevant data was acquired. Initially, it is necessary to define the scientific database. Although there are a multitude of bibliographic databases (e.g., SpringerLink, PubMed, arXiv, etc.), not all provide the essential information required to employ the aforementioned techniques effectively. In addition to adhering to the quality criteria of the listed publications, this analysis requires access to the abstract and citation information. With the discontinuation of Microsoft Academics, the predominant databases for bibliographic analysis include Web of Science (WoS), Scopus, Google Scholar, and Dimensions \cite{Moral.2020}. In particular, Google Scholar entries are not downloadable, and preliminary evaluations indicate that Dimensions lacks complete database entries (e.g., missing the abstract or publication channel). The Web of Science Core Collection fulfils all the specified criteria. Given that the WoS criteria exclude a significant number of conferences, which serve as a vital publication channel in the broader modelling field, Scopus was also selected as a supplementary database. Thus, whenever possible, a combination of both datasets was utilised to expand the research base. The procedures for harmonization are elaborated upon in the following section.\\
\begin{figure*}[t!]
\centering
\subfloat[]{
\resizebox{0.48\linewidth}{!}{% Created by tikzDevice version 0.12.6 on 2025-07-19 14:41:43
% !TEX encoding = UTF-8 Unicode
\begin{tikzpicture}[x=1pt,y=1pt]
\definecolor{fillColor}{RGB}{255,255,255}
\path[use as bounding box,fill=fillColor,fill opacity=0.00] (0,0) rectangle (264.51,180.67);
\begin{scope}
\path[clip] ( 18.94, 18.12) rectangle (264.51,156.76);
\definecolor{drawColor}{gray}{0.80}

\path[draw=drawColor,line width= 0.2pt,line join=round] ( 18.94, 38.43) --
	(264.51, 38.43);

\path[draw=drawColor,line width= 0.2pt,line join=round] ( 18.94, 69.55) --
	(264.51, 69.55);

\path[draw=drawColor,line width= 0.2pt,line join=round] ( 18.94,100.67) --
	(264.51,100.67);

\path[draw=drawColor,line width= 0.2pt,line join=round] ( 18.94,131.79) --
	(264.51,131.79);

\path[draw=drawColor,line width= 0.2pt,line join=round] ( 45.72, 18.12) --
	( 45.72,156.76);

\path[draw=drawColor,line width= 0.2pt,line join=round] ( 95.22, 18.12) --
	( 95.22,156.76);

\path[draw=drawColor,line width= 0.2pt,line join=round] (144.72, 18.12) --
	(144.72,156.76);

\path[draw=drawColor,line width= 0.2pt,line join=round] (194.22, 18.12) --
	(194.22,156.76);

\path[draw=drawColor,line width= 0.2pt,line join=round] (243.72, 18.12) --
	(243.72,156.76);

\path[draw=drawColor,line width= 0.2pt,line join=round] ( 18.94, 22.87) --
	(264.51, 22.87);

\path[draw=drawColor,line width= 0.2pt,line join=round] ( 18.94, 53.99) --
	(264.51, 53.99);

\path[draw=drawColor,line width= 0.2pt,line join=round] ( 18.94, 85.11) --
	(264.51, 85.11);

\path[draw=drawColor,line width= 0.2pt,line join=round] ( 18.94,116.23) --
	(264.51,116.23);

\path[draw=drawColor,line width= 0.2pt,line join=round] ( 18.94,147.35) --
	(264.51,147.35);

\path[draw=drawColor,line width= 0.2pt,line join=round] ( 20.97, 18.12) --
	( 20.97,156.76);

\path[draw=drawColor,line width= 0.2pt,line join=round] ( 70.47, 18.12) --
	( 70.47,156.76);

\path[draw=drawColor,line width= 0.2pt,line join=round] (119.97, 18.12) --
	(119.97,156.76);

\path[draw=drawColor,line width= 0.2pt,line join=round] (169.47, 18.12) --
	(169.47,156.76);

\path[draw=drawColor,line width= 0.2pt,line join=round] (218.97, 18.12) --
	(218.97,156.76);
\definecolor{drawColor}{RGB}{216,183,10}

\path[draw=drawColor,line width= 1.7pt,line join=round] ( 20.97, 24.43) --
	( 90.27, 25.98) --
	(119.97, 24.43) --
	(129.87, 27.54) --
	(134.82, 24.43) --
	(144.72, 24.43) --
	(149.67, 25.98) --
	(154.62, 24.43) --
	(159.57, 33.76) --
	(164.52, 35.32) --
	(169.47, 39.99) --
	(174.42, 38.43) --
	(179.37, 52.43) --
	(184.32, 39.99) --
	(189.27, 47.77) --
	(194.22, 74.22) --
	(199.17, 82.00) --
	(204.12, 86.67) --
	(209.07,130.23) --
	(214.02,150.46) --
	(218.97,139.57);
\definecolor{drawColor}{RGB}{162,164,117}

\path[draw=drawColor,line width= 1.7pt,line join=round] (194.22, 24.43) --
	(199.17, 24.43) --
	(204.12, 24.43) --
	(209.07, 25.98) --
	(214.02, 24.43);
\definecolor{drawColor}{RGB}{151,45,21}

\path[draw=drawColor,line width= 1.7pt,line join=round] ( 20.97, 24.43) --
	( 25.92, 24.43) --
	( 65.52, 25.98) --
	( 90.27, 24.43) --
	(110.07, 25.98) --
	(119.97, 27.54) --
	(129.87, 24.43) --
	(134.82, 25.98) --
	(139.77, 25.98) --
	(144.72, 32.21) --
	(149.67, 35.32) --
	(154.62, 44.65) --
	(159.57, 30.65) --
	(164.52, 49.32) --
	(169.47, 46.21) --
	(174.42, 39.99) --
	(179.37, 43.10) --
	(184.32, 36.87) --
	(189.27, 58.66) --
	(194.22, 60.21) --
	(199.17,102.23) --
	(204.12, 77.33) --
	(209.07, 82.00) --
	(214.02,100.67) --
	(218.97, 88.22);
\definecolor{drawColor}{RGB}{0,0,0}

\node[text=drawColor,anchor=base west,inner sep=0pt, outer sep=0pt, scale=  0.57] at (218.97, 29.93) {BOOK};

\node[text=drawColor,anchor=base west,inner sep=0pt, outer sep=0pt, scale=  0.57] at (218.97, 21.74) {CHAPTER};

\node[text=drawColor,anchor=base west,inner sep=0pt, outer sep=0pt, scale=  0.57] at (223.92,137.61) {ARTICLE};

\node[text=drawColor,anchor=base west,inner sep=0pt, outer sep=0pt, scale=  0.57] at (220.20, 90.36) {CONFERENCE};

\node[text=drawColor,anchor=base west,inner sep=0pt, outer sep=0pt, scale=  0.57] at (220.20, 82.17) {PAPER};
\end{scope}
\begin{scope}
\path[clip] (  0.00,  0.00) rectangle (264.51,180.67);
\definecolor{drawColor}{gray}{0.30}

\node[text=drawColor,anchor=base east,inner sep=0pt, outer sep=0pt, scale=  0.70] at ( 16.06, 20.46) {0};

\node[text=drawColor,anchor=base east,inner sep=0pt, outer sep=0pt, scale=  0.70] at ( 16.06, 51.58) {20};

\node[text=drawColor,anchor=base east,inner sep=0pt, outer sep=0pt, scale=  0.70] at ( 16.06, 82.70) {40};

\node[text=drawColor,anchor=base east,inner sep=0pt, outer sep=0pt, scale=  0.70] at ( 16.06,113.82) {60};

\node[text=drawColor,anchor=base east,inner sep=0pt, outer sep=0pt, scale=  0.70] at ( 16.06,144.94) {80};
\end{scope}
\begin{scope}
\path[clip] (  0.00,  0.00) rectangle (264.51,180.67);
\definecolor{drawColor}{gray}{0.30}

\node[text=drawColor,anchor=base,inner sep=0pt, outer sep=0pt, scale=  0.70] at ( 20.97, 10.43) {1984};

\node[text=drawColor,anchor=base,inner sep=0pt, outer sep=0pt, scale=  0.70] at ( 70.47, 10.43) {1994};

\node[text=drawColor,anchor=base,inner sep=0pt, outer sep=0pt, scale=  0.70] at (119.97, 10.43) {2004};

\node[text=drawColor,anchor=base,inner sep=0pt, outer sep=0pt, scale=  0.70] at (169.47, 10.43) {2014};

\node[text=drawColor,anchor=base,inner sep=0pt, outer sep=0pt, scale=  0.70] at (218.97, 10.43) {2024};
\end{scope}
\begin{scope}
\path[clip] (  0.00,  0.00) rectangle (264.51,180.67);
\definecolor{drawColor}{RGB}{0,0,0}

\node[text=drawColor,anchor=base east,inner sep=0pt, outer sep=0pt, scale=  0.70] at (264.51,  1.36) {\bfseries Year};
\end{scope}
\begin{scope}
\path[clip] (  0.00,  0.00) rectangle (264.51,180.67);
\definecolor{drawColor}{RGB}{0,0,0}

\node[text=drawColor,rotate= 90.00,anchor=base east,inner sep=0pt, outer sep=0pt, scale=  0.70] at (  4.83,156.76) {\bfseries Publications};
\end{scope}
\begin{scope}
\path[clip] (  0.00,  0.00) rectangle (264.51,180.67);
\definecolor{drawColor}{RGB}{0,0,0}

\node[text=drawColor,anchor=base west,inner sep=0pt, outer sep=0pt, scale=  0.80] at ( 18.94,163.32) {Combined WoS and Scopus data};
\end{scope}
\begin{scope}
\path[clip] (  0.00,  0.00) rectangle (264.51,180.67);
\definecolor{drawColor}{RGB}{0,0,0}

\node[text=drawColor,anchor=base west,inner sep=0pt, outer sep=0pt, scale=  1.00] at ( 18.94,173.77) {\bfseries Annual Scientific Production by Publication Type};
\end{scope}
\end{tikzpicture}}
\label{fig:prod}
}
\hfil
\subfloat[]{
    \resizebox{0.48\textwidth}{!}{\input{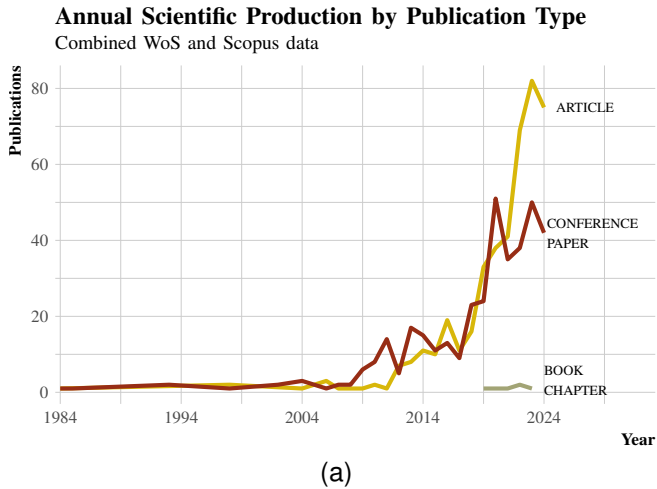}
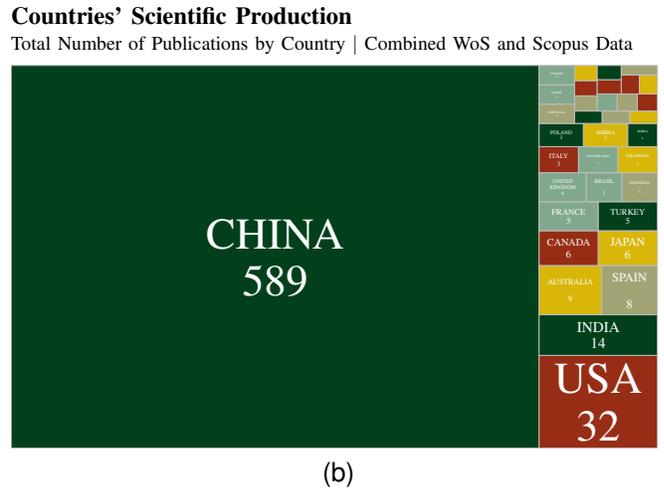}%
    \label{fig:cou}
}
\caption{Scientific production analysis. (a) Productivity of research constituents by publication type. Until ~2008 a small total number of publications were published. Between 2008 and 2017, the type of publications increased, whereas the publication type remained comparable. From 2017 until today, the total number has increased faster with a higher number of articles published. Furthermore, the first book chapters appeared. (b) Country of scientific production based on the corresponding author. With 72\%, China is at a distance the country with the highest scientific production, followed by the USA with 4\% and India with 2\%. It has to be noted that 11.1\% of the corresponding authors are missing, which leads to a skewed depiction.}
\label{fig:science_overview}
\end{figure*}
Subsequently, established terms from the literature were recorded, collectively reviewed, and expanded by the authors. Both inclusion and exclusion criteria were defined. Ten query variants were tested to ascertain their suitability and to evaluate the breadth of the query, with evaluations based on a sample using the publication title. The assessment criterion focused on thematic alignment, specifically the degree to which a publication is related to the area of modelling passenger flows for local public transit. For example, broader queries often encompassed unrelated modes of transport, such as automobiles or airplanes, while overly narrow queries omitted certain relevant areas. In conjunction with the search terms, criteria such as English language, publications up to the end of 2024, terms in the title, and publication types, articles, proceedings, and chapters were defined as inclusion criteria. Upon evaluation, the definitive search query is composed of four distinct components. The initial component encompasses diverse combinations of word pairs pertinent to individuals in local public transit (e.g., ``passenger flow*", ``passenger load*", etc.). The second component encapsulates the forecasting or modelling dimension (e.g., ``predict*", ``model*", etc.). The third component delineates the operational domain involving buses and rail-based transit modes (``bus*", ``train", ``tram*" etc.). The fourth and final component establishes exclusion criteria (NOT ``airplane", ``ferry", etc.) to systematically eliminate unrelated transport modes. Consequently, the formulated search query is as follows:
\\\newline
\textit{((``passenger load*" OR ``passenger flow*" OR ``passenger demand*" OR ``passenger volum*" OR ``ridership'' OR ``passenger occupancy" OR ``passenger capacity" OR ``passenger density") AND (predict* OR forecast* OR simulat* or estimat* OR model*) AND (bus* OR train OR trains OR ``urban rail transit" OR metro OR tube OR subway OR rail* OR tram*) NOT (airplane OR airport OR airline or ferry or taxi))}
\subsection{Run analysis}
\noindent The primary components of the analysis were performed using the R package \textit{bibliometrix} \cite{Aria.2017}, due to its high flexibility and inclusion of most analyses relevant to our research \cite{Moral.2020}. The relevant implementation details are described below.

The documents listed until 01.03.2025 in Scopus and WoS were retrieved. Subsequent to retrieval, a replacement of the full name was performed where feasible, and documents $>=$ 2025 were filtered. WoS and Scopus datasets were merged, and duplicates were removed. In addition to duplicate elimination, document types were standardised, for instance, converting `proceedings paper' to `conference paper.' Standardization was also applied to publishing institutions, as inconsistencies in spelling and abbreviations were observed. For example, `university' was unified as `univ,' and `of' was omitted where applicable. Given the manual data cleaning process, the occurrence of minor errors is expected. The quality of the preprocessed data was subsequently verified through descriptive measures and manual review. Afterwards, the actual techniques were performed and visualised.

Most analytical techniques are self-explanatory, with the exception of a particular implementation of a directed co-citation network variant. The objective was to integrate the co-citation network of "passenger flow" within the broader modelling field. To achieve this, two additional queries were executed that focus on ML and time series. The three resulting data sets were merged to construct a co-citation network incorporating information about the original queries. The spatial distinction between the `Time series' and `ML' networks and the `Transit' network was then accentuated to highlight their connections. Thus, a visual representation of the higher-level trends was achieved.

\section{Results}
\subsection{Pre-Processing and Quality Assessment}
\noindent The initial query on WoS yielded 508 documents, while Scopus produced 784 documents. When the results from both databases were merged and duplicates removed, a total of 821 documents were identified. For a detailed analysis of the data quality per category, one can refer to the Table~\ref{tab:missing_data}. In general, the relevant categories are available at an adequate level of quality. Subsequently, documents classified under erroneous typologies such as `NA', `MEETING ABSTRACT', and `CORRECTION' were excluded, resulting in a final reduction to 814 documents. As a further preparatory measure, the names of the affiliations were standardised. In addition to manual verification, the consolidation from 780 to 690 unique corresponding author affiliations illustrates this standardization. The case of Wuhan provides an example of existing limitations. The term 'Wuhan*' appears in various affiliations, including ‘WUHAN COMPREHENSIVE TRANSPORTRESEARCH INST CO’, ‘WUHAN RAILWAY VOCATIONAL COLLEGE TECHNOL’, ‘WUHAN UNIV’, ‘WUHAN UNIV SCI AND TECHNOL', and 'WUHAN UNIV TECHNOL'. The degree to which these names are consolidated is beyond the scope of this study.

\subsection{Content-related Results}

% \begin{figure}[t!]
% \centering
% \resizebox{\columnwidth}{!}{\input{fig/rel_sources_viv_yg}}
% \caption{Top ten sources by number of publications. It becomes clear that there is no main publication channel. The figure does not sufficiently illustrate the importance of conferences, as these are each evaluated individually. As an example, the International Conference of Transportation Professionals (CICTP) has cumulatively contributed 35 publications throughout the years, with a peak of seven publications in 2019.}
% \label{fig:sou}
% \end{figure}

\noindent As already mentioned, a total of 814 eligible documents were available during the publication period from 1984 to 2024. Fig. \ref{fig:prod} illustrates that while sporadic publications occurred before 2008, the domain of passenger flow prediction in local public transport has shown a noticeable growth only from 2008 onward, with a significantly increased number of articles published from 2017 to the present. Concerning the types of publication, there was an equilibrium between conference papers and journal articles until approximately 2020, after which a shift toward journal articles has been observed. It remains unclear whether this shift indicates a genuine trend or is simply a temporary effect of the COVID-19 pandemic. Further document types are of minor occurrence.
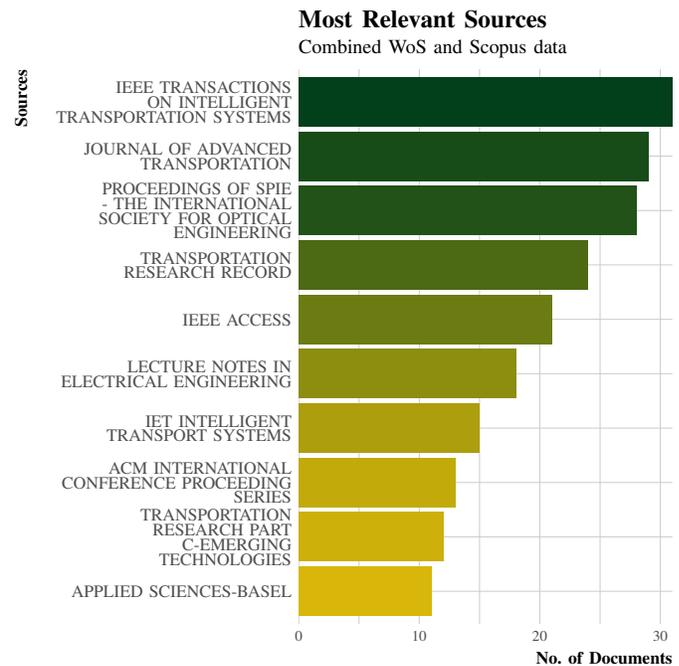
\begin{figure}[b!]
\centering
\resizebox{\columnwidth}{!}{% Created by tikzDevice version 0.12.6 on 2025-10-02 11:25:26
% !TEX encoding = UTF-8 Unicode
\begin{tikzpicture}[x=1pt,y=1pt]
\definecolor{fillColor}{RGB}{255,255,255}
\path[use as bounding box,fill=fillColor,fill opacity=0.00] (0,0) rectangle (264.51,264.51);
\begin{scope}
\path[clip] (114.19, 17.24) rectangle (264.51,240.60);
\definecolor{drawColor}{gray}{0.80}

\path[draw=drawColor,line width= 0.2pt,line join=round] (138.44, 17.24) --
	(138.44,240.60);

\path[draw=drawColor,line width= 0.2pt,line join=round] (186.93, 17.24) --
	(186.93,240.60);

\path[draw=drawColor,line width= 0.2pt,line join=round] (235.42, 17.24) --
	(235.42,240.60);

\path[draw=drawColor,line width= 0.2pt,line join=round] (114.19, 30.38) --
	(264.51, 30.38);

\path[draw=drawColor,line width= 0.2pt,line join=round] (114.19, 52.28) --
	(264.51, 52.28);

\path[draw=drawColor,line width= 0.2pt,line join=round] (114.19, 74.17) --
	(264.51, 74.17);

\path[draw=drawColor,line width= 0.2pt,line join=round] (114.19, 96.07) --
	(264.51, 96.07);

\path[draw=drawColor,line width= 0.2pt,line join=round] (114.19,117.97) --
	(264.51,117.97);

\path[draw=drawColor,line width= 0.2pt,line join=round] (114.19,139.87) --
	(264.51,139.87);

\path[draw=drawColor,line width= 0.2pt,line join=round] (114.19,161.77) --
	(264.51,161.77);

\path[draw=drawColor,line width= 0.2pt,line join=round] (114.19,183.66) --
	(264.51,183.66);

\path[draw=drawColor,line width= 0.2pt,line join=round] (114.19,205.56) --
	(264.51,205.56);

\path[draw=drawColor,line width= 0.2pt,line join=round] (114.19,227.46) --
	(264.51,227.46);

\path[draw=drawColor,line width= 0.2pt,line join=round] (114.19, 17.24) --
	(114.19,240.60);

\path[draw=drawColor,line width= 0.2pt,line join=round] (162.68, 17.24) --
	(162.68,240.60);

\path[draw=drawColor,line width= 0.2pt,line join=round] (211.17, 17.24) --
	(211.17,240.60);

\path[draw=drawColor,line width= 0.2pt,line join=round] (259.66, 17.24) --
	(259.66,240.60);
\definecolor{fillColor}{RGB}{216,183,10}

\path[fill=fillColor] (114.19, 20.52) rectangle (167.53, 40.23);
\definecolor{fillColor}{RGB}{205,176,10}

\path[fill=fillColor] (114.19, 42.42) rectangle (172.38, 62.13);
\definecolor{fillColor}{RGB}{194,170,11}

\path[fill=fillColor] (114.19, 64.32) rectangle (177.23, 84.03);
\definecolor{fillColor}{RGB}{172,158,13}

\path[fill=fillColor] (114.19, 86.22) rectangle (186.93,105.93);
\definecolor{fillColor}{RGB}{141,141,16}

\path[fill=fillColor] (114.19,108.12) rectangle (201.47,127.82);
\definecolor{fillColor}{RGB}{108,123,18}

\path[fill=fillColor] (114.19,130.01) rectangle (216.02,149.72);
\definecolor{fillColor}{RGB}{76,105,20}

\path[fill=fillColor] (114.19,151.91) rectangle (230.57,171.62);
\definecolor{fillColor}{RGB}{34,82,24}

\path[fill=fillColor] (114.19,173.81) rectangle (249.96,193.52);
\definecolor{fillColor}{RGB}{23,76,25}

\path[fill=fillColor] (114.19,195.71) rectangle (254.81,215.42);
\definecolor{fillColor}{RGB}{2,64,27}

\path[fill=fillColor] (114.19,217.61) rectangle (264.51,237.31);
\end{scope}
\begin{scope}
\path[clip] (  0.00,  0.00) rectangle (264.51,264.51);
\definecolor{drawColor}{gray}{0.30}

\node[text=drawColor,anchor=base east,inner sep=0pt, outer sep=0pt, scale=  0.70] at (111.32, 27.97) {APPLIED SCIENCES-BASEL};

\node[text=drawColor,anchor=base east,inner sep=0pt, outer sep=0pt, scale=  0.70] at (111.32, 58.69) {TRANSPORTATION};

\node[text=drawColor,anchor=base east,inner sep=0pt, outer sep=0pt, scale=  0.70] at (111.32, 52.81) {RESEARCH PART};

\node[text=drawColor,anchor=base east,inner sep=0pt, outer sep=0pt, scale=  0.70] at (111.32, 46.93) {C-EMERGING};

\node[text=drawColor,anchor=base east,inner sep=0pt, outer sep=0pt, scale=  0.70] at (111.32, 41.05) {TECHNOLOGIES};

\node[text=drawColor,anchor=base east,inner sep=0pt, outer sep=0pt, scale=  0.70] at (111.32, 77.64) {ACM INTERNATIONAL};

\node[text=drawColor,anchor=base east,inner sep=0pt, outer sep=0pt, scale=  0.70] at (111.32, 71.76) {CONFERENCE PROCEEDING};

\node[text=drawColor,anchor=base east,inner sep=0pt, outer sep=0pt, scale=  0.70] at (111.32, 65.88) {SERIES};

\node[text=drawColor,anchor=base east,inner sep=0pt, outer sep=0pt, scale=  0.70] at (111.32, 96.60) {IET INTELLIGENT};

\node[text=drawColor,anchor=base east,inner sep=0pt, outer sep=0pt, scale=  0.70] at (111.32, 90.72) {TRANSPORT SYSTEMS};

\node[text=drawColor,anchor=base east,inner sep=0pt, outer sep=0pt, scale=  0.70] at (111.32,118.50) {LECTURE NOTES IN};

\node[text=drawColor,anchor=base east,inner sep=0pt, outer sep=0pt, scale=  0.70] at (111.32,112.62) {ELECTRICAL ENGINEERING};

\node[text=drawColor,anchor=base east,inner sep=0pt, outer sep=0pt, scale=  0.70] at (111.32,137.46) {IEEE ACCESS};

\node[text=drawColor,anchor=base east,inner sep=0pt, outer sep=0pt, scale=  0.70] at (111.32,162.30) {TRANSPORTATION};

\node[text=drawColor,anchor=base east,inner sep=0pt, outer sep=0pt, scale=  0.70] at (111.32,156.42) {RESEARCH RECORD};

\node[text=drawColor,anchor=base east,inner sep=0pt, outer sep=0pt, scale=  0.70] at (111.32,190.07) {PROCEEDINGS OF SPIE};

\node[text=drawColor,anchor=base east,inner sep=0pt, outer sep=0pt, scale=  0.70] at (111.32,184.19) {- THE INTERNATIONAL};

\node[text=drawColor,anchor=base east,inner sep=0pt, outer sep=0pt, scale=  0.70] at (111.32,178.31) {SOCIETY FOR OPTICAL};

\node[text=drawColor,anchor=base east,inner sep=0pt, outer sep=0pt, scale=  0.70] at (111.32,172.43) {ENGINEERING};

\node[text=drawColor,anchor=base east,inner sep=0pt, outer sep=0pt, scale=  0.70] at (111.32,206.09) {JOURNAL OF ADVANCED};

\node[text=drawColor,anchor=base east,inner sep=0pt, outer sep=0pt, scale=  0.70] at (111.32,200.21) {TRANSPORTATION};

\node[text=drawColor,anchor=base east,inner sep=0pt, outer sep=0pt, scale=  0.70] at (111.32,230.93) {IEEE TRANSACTIONS};

\node[text=drawColor,anchor=base east,inner sep=0pt, outer sep=0pt, scale=  0.70] at (111.32,225.05) {ON INTELLIGENT};

\node[text=drawColor,anchor=base east,inner sep=0pt, outer sep=0pt, scale=  0.70] at (111.32,219.17) {TRANSPORTATION SYSTEMS};
\end{scope}
\begin{scope}
\path[clip] (  0.00,  0.00) rectangle (264.51,264.51);
\definecolor{drawColor}{gray}{0.30}

\node[text=drawColor,anchor=base,inner sep=0pt, outer sep=0pt, scale=  0.60] at (114.19, 10.23) {0};

\node[text=drawColor,anchor=base,inner sep=0pt, outer sep=0pt, scale=  0.60] at (162.68, 10.23) {10};

\node[text=drawColor,anchor=base,inner sep=0pt, outer sep=0pt, scale=  0.60] at (211.17, 10.23) {20};

\node[text=drawColor,anchor=base,inner sep=0pt, outer sep=0pt, scale=  0.60] at (259.66, 10.23) {30};
\end{scope}
\begin{scope}
\path[clip] (  0.00,  0.00) rectangle (264.51,264.51);
\definecolor{drawColor}{RGB}{0,0,0}

\node[text=drawColor,anchor=base east,inner sep=0pt, outer sep=0pt, scale=  0.70] at (264.51,  1.36) {\bfseries No. of Documents};
\end{scope}
\begin{scope}
\path[clip] (  0.00,  0.00) rectangle (264.51,264.51);
\definecolor{drawColor}{RGB}{0,0,0}

\node[text=drawColor,rotate= 90.00,anchor=base east,inner sep=0pt, outer sep=0pt, scale=  0.70] at (  4.83,240.60) {\bfseries Sources};
\end{scope}
\begin{scope}
\path[clip] (  0.00,  0.00) rectangle (264.51,264.51);
\definecolor{drawColor}{RGB}{0,0,0}

\node[text=drawColor,anchor=base west,inner sep=0pt, outer sep=0pt, scale=  0.80] at (114.19,247.15) {Combined WoS and Scopus data};
\end{scope}
\begin{scope}
\path[clip] (  0.00,  0.00) rectangle (264.51,264.51);
\definecolor{drawColor}{RGB}{0,0,0}

\node[text=drawColor,anchor=base west,inner sep=0pt, outer sep=0pt, scale=  1.00] at (114.19,257.61) {\bfseries Most Relevant Sources};
\end{scope}
\end{tikzpicture}}
\caption{Top ten sources by number of publications. It becomes clear that there is no main publication channel. The figure does not sufficiently illustrate the importance of conferences, as these are each evaluated individually. As an example, the International Conference of Transportation Professionals (CICTP) has cumulatively contributed 35 publications throughout the years, with a peak of seven publications in 2019.}
\label{fig:sou}
\end{figure}
Among the 814 documents analysed, 72\% were produced with a Chinese corresponding author, suggesting a possible bias due to the exclusion of publications in Chinese (see Fig. \ref{fig:cou}). This is supported by the identification of 129 additional publications in Chinese within the Scopus database. Consequently, the main publishing institutions are various transportation-focused universities, including Beijing Jiaotong University, Lanzhou Jiaotong University, Southeast University, Southwest Jiaotong University, and Tongji University. Beijing Jiaotong University is notably prominent in terms of publication volume, and other mentioned universities exhibit similar publishing patterns (see Fig. \ref{fig:pub_inst}). The first institutions outside China to appear in this context are the City University of Hong Kong and the University of São Paulo, which rank 11th and 40th, respectively.

The ten most frequent publication channels depicted in Fig. \ref{fig:sou} comprise 25\% of all identified channels, with a total of 415 distinct channels found. This substantial number of publication channels indicates a high degree of interdisciplinarity and an evolving field. This observation is corroborated by the fact that only 27 publication channels have published $>=5$ documents. However, it is important to acknowledge that conferences are evaluated individually, which results in their significance being underrepresented in Fig. \ref{fig:sou} and an overrepresentation of the diversity of publication channels. For example, the International Conference of Transportation Professionals (CICTP) has cumulatively contributed 35 publications throughout the years, with a peak of seven publications in 2019.

\begin{figure*}[b]
\centering
\resizebox{\textwidth}{!}{\input{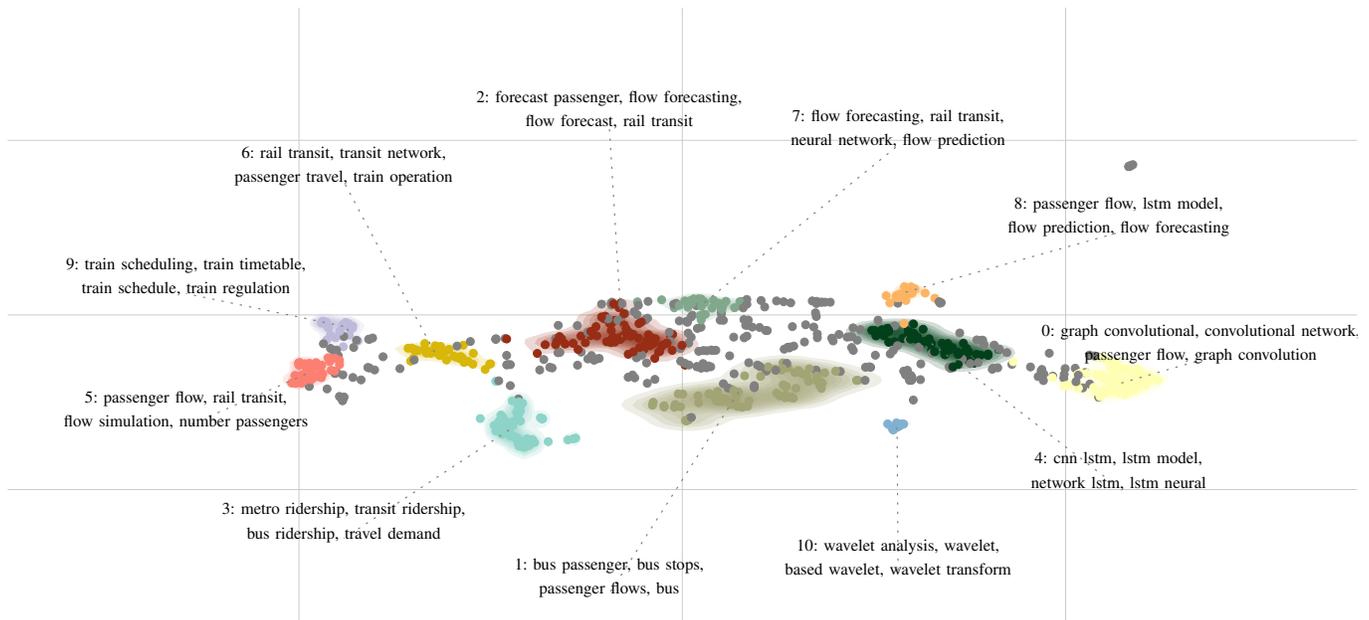}}
\caption{Utilizing the embeddings of abstracts generated through the ``MiniLM-L6-v2" model within the SentenceTransformers framework, eleven distinct clusters have been identified. Within the documents associated with a specific cluster, some documents lack a clear association with any cluster and are thus categorised as ambiguous (grey). The topic representation, facilitated by the "KeyBERTInspired function," illustrates various modes and used methods.}
\label{fig:bert}
\end{figure*}

Beyond qualitative attributes, the authors seek to evaluate the scope or potential significance within a field. Notably, in terms of single document citations, 'Transportation Research Part C: Emerging Technologies' and 'Neurocomputing' are distinguished as publication channels with extensive reach (refer to Table~\ref{tab:pub_channel}). Within the publication volume ranking, 'Transportation Research Part C: Emerging Technologies' occupies the ninth position, with twelve published articles.

To summarize the performance analysis, it can be said that the importance of the field has grown steadily since around 2008 and is expected to continue to increase. The combination of technical modelling components and the breadth of content in the field is reflected in the breadth of publication channels.

% \begin{figure*}[t!]
% \centering
% \resizebox{\textwidth}{!}{\input{fig/bertopic_keybert_minillm.tex}}
% \caption{Utilizing the embeddings of abstracts generated through the ``MiniLM-L6-v2" model within the SentenceTransformers framework, eleven distinct clusters have been identified. Within the documents associated with a specific cluster, some documents lack a clear association with any cluster and are thus categorised as ambiguous (grey). The topic representation, facilitated by the "KeyBERTInspired function," illustrates various modes and used methods.}
% \label{fig:bert}
% \end{figure*}
To illustrate the developments and relationships for science mapping, we can start with the self-perception of authors within the field. An indicator of this is the selection of keywords by the authors, which can yield insights regarding the intended or actual target group. In addition, the use of rather standardised keywords can indicate a common understanding within a field. In this analysis, the keywords used most frequently per article are `PASSENGER FLOW PREDICTION (77)', `URBAN RAIL TRANSIT (74)', `PASSENGER FLOW (69)', `DEEP LEARNING (65)', and `FORECASTING (29)'. Against the background of a total of 814 documents, the most frequent keyword is used in 9.5\% and the fifth most frequent in 3.5\% of cases. These values indicate that there is no distinct understanding as a field or that the field analyzed here could have further subfields that are more closed. All keywords fit the field in terms of content, although the employment of the keyword 'DEEP LEARNING' appears to be directed towards engaging a broader audience.

Beyond facilitating self-understanding, topic modelling based on abstracts offers deeper insights into thematic clusters (see Fig. \ref{fig:bert}). Among the 11 identified clusters, four on the right side (clusters 0, 4, 8, and 10) are primarily associated with specific modelling techniques. The smallest cluster, cluster 10, includes the term wavelet transform, which is frequently used as a preprocessing step for statistical models such as autoregressive models or in the context of combined models \cite{Xue.2023}. When additionally examining the grouped trigrams presented in Fig. \ref{fig:trigram}, ARIMA, as well as Empirical Mode Decomposition, are identified as additional methods. The remaining three clusters are associated with deep learning (DL) approaches. Clusters 4 and 8 refer to specific types of neural networks, namely long- and short-term memory (LSTM) and convolutional neural networks (CNN), while Cluster 0 highlights the use of graph convolutional networks (GCN). Complementary methods, such as support vector machines (SVM), also emerge from the trigram analysis.

A comparative analysis of the other clusters reveals a general pattern of focus. For example, the term bus appears in only two clusters, whereas rail-based transit is represented in six clusters, underscoring the dominance of studies related to rail transport. Clusters 5 and 9 appear to focus on the management of passenger flows within stations and broader transit operations, as indicated by terms such as flow simulation and train regulation. Phrases such as transit network and train operation further suggest an emphasis on network- or system-level management strategies. It is also worth noting that
cluster 2 contains the term most frequently used, passenger flow forecast, which probably represents the overarching theme of many documents. Several documents could not be clearly assigned to a single cluster, likely due to the multidimensional nature of their content or potential distortions introduced by dimensionality reduction techniques.

In general, some important themes identified in the trigram analysis (Fig. \ref{fig:trigram}) are not reflected directly in the topic groups. These include the nature of the data sources, which most commonly consist of historical data and/or automatic fare collection (AFC) data. Furthermore, no specific type of traffic flow is distinctly represented in the clusters. The trigram analysis suggests that all types of flow are addressed with comparable frequency.
\begin{figure*}[b]
    \centering
    \resizebox{\textwidth}{!}{\input{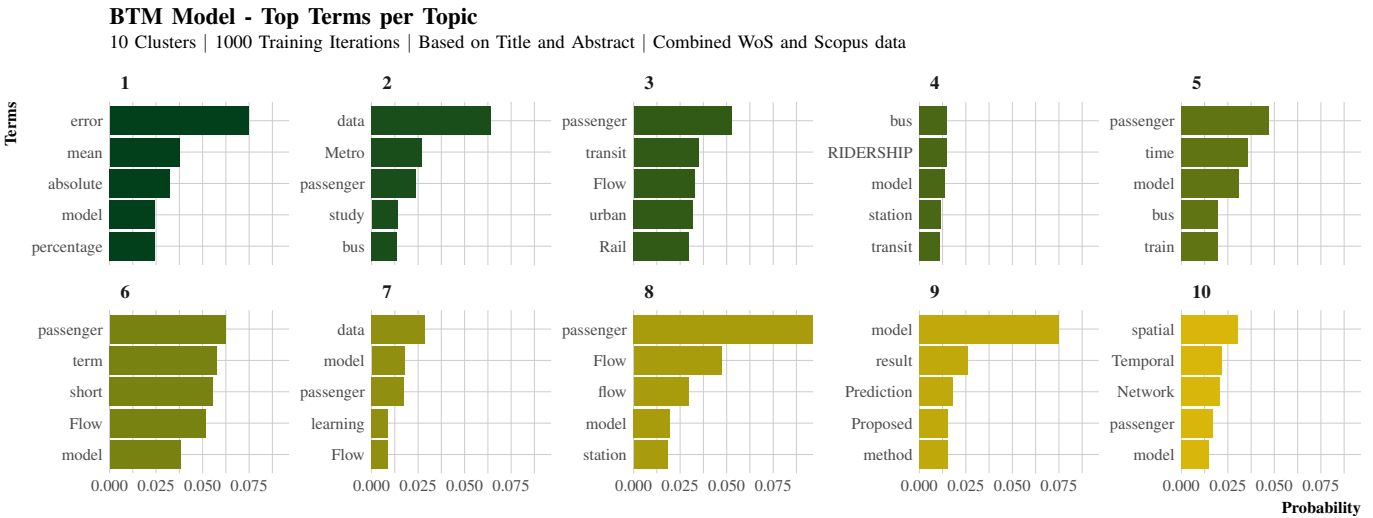}}
    \caption{Clustered co-occurrence patterns via a Biterm Topic Model (BTM). Each cluster is related to some content topic. A higher percentage depicts a higher probability for such a word combination in the corpus of documents.}
    \label{fig:enter-label}
\end{figure*}

Next to grouping the single documents into clusters, Biterm Topic Modelling (BTM) can be employed as a form of co-word analysis to explore additional themes through co-occurrence patterns across entire documents (Fig. \ref{fig:enter-label}). The number of ten individual clusters was determined by iterative approximation, with the title and abstract combined as input text. Beyond the already identified findings that are not revisited here, such as the association of cluster 6 with neural networks or LSTM models, additional content-related correlations can be discerned. Cluster 1 is associated with common evaluation metrics, including mean squared error, root mean square error, and mean absolute percentage error. Cluster 3 underscores the integration of urban rail transit (URT) with its management and operations. Lastly, Cluster 2 highlights that empirical studies that utilize real-world data are frequently conducted in China, which aligns with the established global predominance.

\begin{figure}[t!]
\centering
\resizebox{\columnwidth}{!}{\input{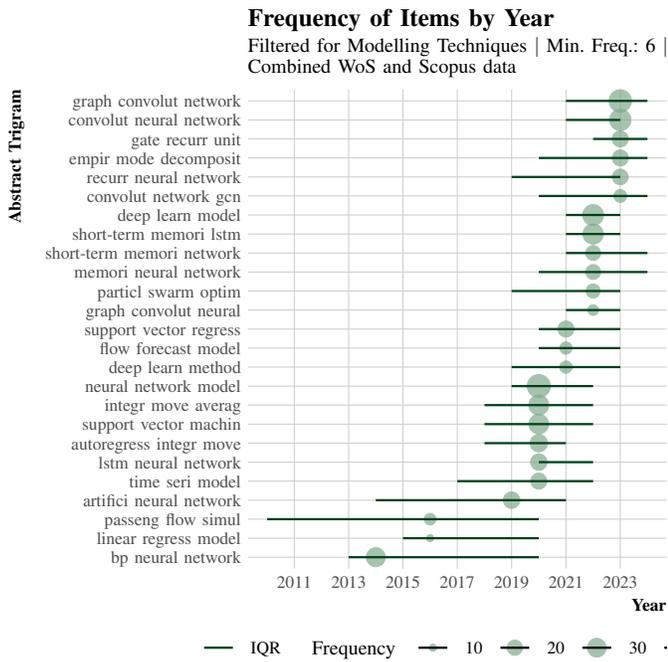}}
\caption{Filtered trigrams in relation to their main usage time frame. Until |
~2018, generally broader modelling terms were frequently used. Between 2018-2023 traditional time series models as well as ML techniques gain relevancy. From 2023-today more specialised deep learning models appeared more frequently.}
\label{fig:fi_trend}
\end{figure}

\begin{figure*}[t!]
\centering
\resizebox{\textwidth}{!}{\input{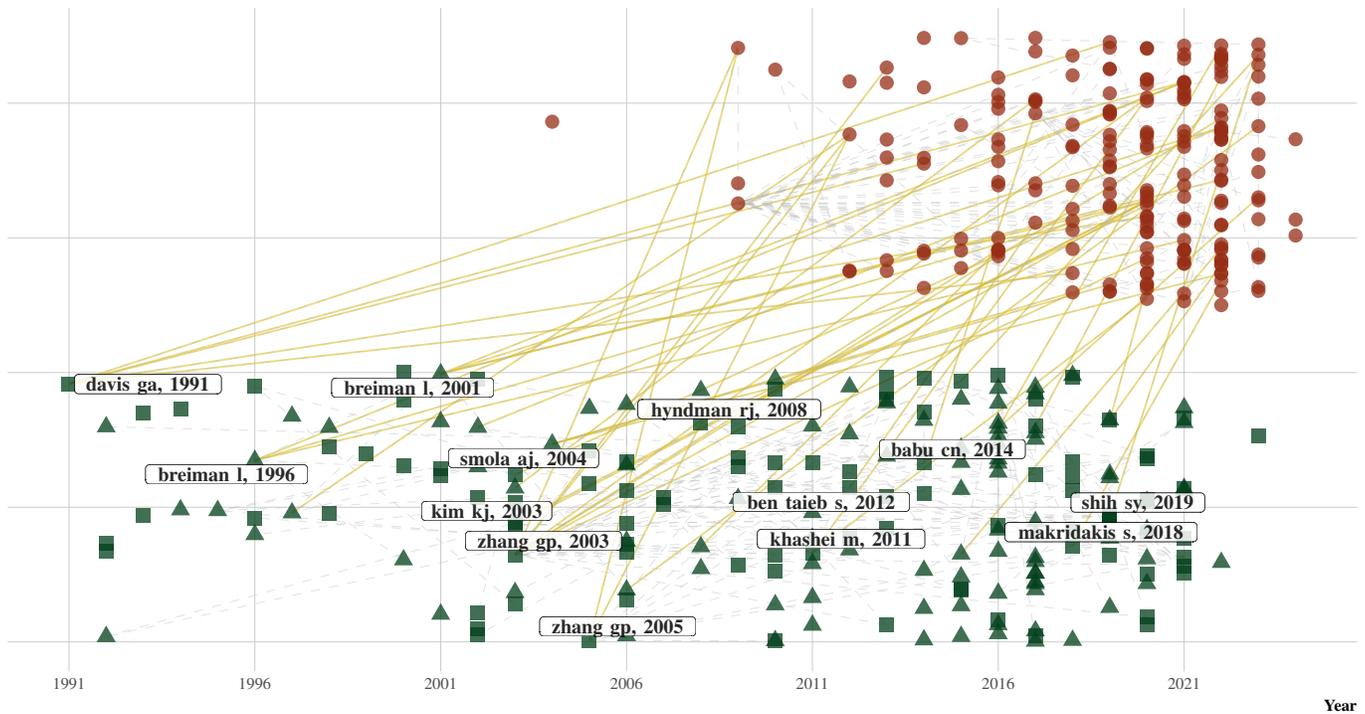}}
\caption{Directed historical citation network based on three different WoS queries. The upper red dots refer to the local transit query and the lower green points depict the time series or deep learning query. Thin grey lines depict connections between queries, and yellow lines cross-thematic connections. Shortenings are provided for the most connected documents from the DL/TS query. Position on the y-axis does not follow a rule.}
\label{fig:comb_hist}
\end{figure*}

In a final analysis, which can be classified as a co-word analysis, the trigrams were initially organised according to their temporal frequency and subsequently filtered according to the modelling approach (see Fig. \ref{fig:fi_trend}). As noted previously, absolute values should be interpreted with caution, but distinct developmental stages can be observed. Trigrams before 2010 are excluded because of the limited number of total publications. Until approximately 2018, broader modelling terms such as "linear regression model" and "artificial neural network" were frequently mentioned, reflecting an early stage dominated by traditional statistical and general machine learning methods. From 2018 onward, there is a marked increase in methodological diversity, which corresponds to the overall increase in publications (Fig. \ref{fig:prod}). During this transition phase (2018–2023), both time series models (e.g., "integrated moving average") and ML techniques (e.g., "support vector machine") gain traction. Notably, while deep learning approaches emerge, they are initially referenced in general terms such as "deep learning model" or "neural network model" rather than specific architectures. From 2023 to the present, there has been a shift toward more specialised deep learning models, including the "convolutional neural network" and the "recurrent neural network." This trend suggests a maturing research landscape in which studies increasingly focus on architecture-specific implementations rather than broad methodological categories.\\
Extending on the development within the field, Fig. \ref{fig:comb_hist} shows connections to the superior fields of ML and general time series modelling. The initially less steep connections between the two fields and therefore the longer adaptation time are clearly visible. In addition, it can be seen that reference has also been made to these older publications in recent years. In recent times, adaptation cycles have been significantly shortened. However, in general, connections are also less frequent in absolute terms.

Upon examining in greater depth the abbreviations of authors frequently associated with this domain, several illustrative examples emerge. In particular, the works of Leo Breiman, which focus on bootstrap aggregation and the concept of random forests, are identifiable as foundational elements of traditional machine learning methodologies \cite{Breiman.1996, Breiman.2001}. Similarly, the contributions of Peter G. Zhang are discernible at the confluence of neural networks and time-series analysis \cite{Zhang.2003, Zhang.2005}. In recent times, advances in deep learning have facilitated the development of more specialised methodologies for time series forecasting. For example, \cite{Shih.2019} proposes a temporal pattern attention mechanism that employs frequency domain representations to extract time-invariant temporal patterns, effectively addressing the challenges associated with modelling long-term dependencies in multivariate datasets. This approach signifies a significant progression from conventional deep learning models towards architectures specifically crafted to accommodate complex temporal dynamics.
\subsection{Synthesis of existing reviews}
\noindent When our findings are juxtaposed and synthesised with those of other studies, it is imperative to acknowledge beforehand that due to the disparate methodologies employed, a direct comparison or assessment of absolute values is not feasible. Instead, the comparison revolves around the general trends.

In the initial section of the results, the meta-information pertaining to the publications is scrutinised. Among the analyzed works, only \cite{Rocco.2024} employed a quantitative approach to their data. This corroborates certain outcomes, such as the distribution of publication types. For example, \cite{Rocco.2024} reports that 53\% of the publications were journal articles while 47\% conference articles, a distribution that is reflected in the bibliometric analysis indicating 53\% articles compared to 46\% conference articles. The extensive variability in the publication channels is also evident. For example, \cite{Rocco.2024} illustrates the distribution of publication channels in both conferences and journals where channels spanning up to 2\% are presumably identified. For conferences, other channels constitute 80\%, and for journals, 48\%. In our analysis, both journals and conferences were collectively considered, with the top ten largest publication channels representing merely 25\% of the total publications.  In contrast to those comparable metrics, no publication has shown the dominance of China in terms of responsible authors.

In terms of content, the synthesis of the four publications with our findings is more complex. Compared with the first secondary publication by \cite{Zhai.2018}, our analysis highlights fewer details and concrete modelling. As an example, AFC data was also identified as the main data source, but specific approaches to acquisition, such as images, infrared sensors, or pressure sensors, are explicitly mentioned in their publication. The superordinate structure with linear, nonlinear, and combined models can also be confirmed for linear and nonlinear models in our study. The combined models appear only indirectly and sporadically in our analysis. For example, in Fig. \ref{fig:trigram}, in which both EMD and LSTM are named, which were combined in \cite{Chen.2019}, for example. With regard to the specific models analyzed, the generic terms such as ANN or SVM are also reflected in the bibliometric analysis, although specific forms such as fuzzy ANN, BP ANN, and multiple-kernel LSSVM cannot be named. In addition to our analysis, grey models and Kalman filters must be added as modelling approaches. Finally, the authors mention that '[\dots] some researchers using deep learning to predict the short-term passenger flow variation'. This development has continued and can certainly be seen as the dominant direction in recent years. The second analysis by \cite{Hussain.2021} sets a different focus, and no contradictions or complementary contexts can be identified.

As already mentioned in the first section, \cite{Xue.2023} first develops a framework in which different types of passenger flow can be categorised. In addition to AFC data, they also work out other auxiliary data, such as weather, air conditions, or social media data, as further possible data sources, which are not recognizable in our analysis and therefore do not yet represent a significant source of data. Unlike \cite{Zhai.2018}, the categorization is made into traditional time series forecasting models (e.g., ARIMA, KF), machine learning models (e.g., SVG, LSTM), and combined models. Thus, a temporal perspective is taken rather than the underlying function of a model.

% \begin{figure}[t!]
% \centering
% \resizebox{\columnwidth}{!}{\input{fig/fi_trend.tex}}
% \caption{Filtered trigrams in relation to their main usage time frame. Until |
% ~2018, generally broader modelling terms were frequently used. Between 2018-2023 traditional time series models as well as ML techniques gain relevancy. From 2023-today more specialised deep learning models appeared more frequently.}
% \label{fig:fi_trend}
% \end{figure}
In addition to the modelling approaches already mentioned here, general regression neural networks, (stacked) autoencoders, generative adversial networks, and transformers are also named. The occurrence of the approaches is not specified. The conclusion emphasizes the importance of multi-data fusion and the fact that it is not cost-efficient to create a separate model for each line, station, and network. As an additional argumentation, \cite{Rocco.2024} agrees with this topic and mentions the focus on improving models as a problem rather than addressing other aspects such as costs, interpretability, privacy, and efficiency. In this context, the problem of limited open and free data sets can also be pointed out.
\section{Discussion}
\noindent In conducting a bibliometric analysis of the field of (short-term) passenger flow modelling in local public transit, a dedicated methodology was employed to provide, for the first time, a comprehensive overview of the subject area. This methodology facilitates not only the identification of temporal trends but also the visualisation of connections to the overarching field.
\subsection{Limitations of the Study}
\noindent However, the selected methodological approach encounters certain limitations, which can be classified into methodological and content-related critiques.

Methodologically, the relatively small sample size of 814 documents deserves attention. While it can be contended that such a sample necessitates considerable manual effort, comparative analyses employ substantially larger datasets (e.g., 12,272 for e-learning \cite{Djeki.2022}, 12,436 for GIScience \cite{Biljecki.2016}, and 1,789 for safety culture \cite{Nunen.2018}). The use of larger data sets facilitates the identification of thematic clusters with greater clarity and certainty, enabling trends to be discerned more reliably and broadly. Consequently, the necessity for filtering, as depicted in Figs. \ref{fig:fi_trend} or \ref{fig:trigram}, might be reduced, thereby enhancing reproducibility. Furthermore, it would be feasible to filter publication channels characterised by a low volume of publications or citation frequencies. Such measures would potentially improve the quality of the analyzed publications and, in turn, enhance their relevance.

Biases should be considered an additional aspect of the methodology. A systematic distortion due to language-based filtering can be presumed, extending beyond literature published in Chinese. Furthermore, WoS and Scopus exclude certain publication channels from other contributors in the field, including studies conducted by transport entities or published on platforms like arXiv.org. Lastly, the restricted operational definition of 'local public transit' warrants attention. Modes of transport such as ferries or cable cars are excluded due to the challenge of distinguishing them from long-distance ferries or ski lifts.

In terms of content, the methodology adopted adopts a comprehensive perspective that provides an overview of the entire field. This approach means that the field, precluding a detailed comparison of the individual modelling approaches, cannot be made, and side streams render less prominent research streams, with few publications and less visibility. Consequently, emerging and innovative methodologies are not readily discernible in this context. Although the work focuses in particular on emphasizing short-term passenger flow, it lacks a clear distinction in the query and analysis with regard to long-term passenger flow, owing to the challenges in operationalization. However, achieving such differentiation would be advantageous.
\subsection{Categorisation of the Findings}
\noindent The field of passenger flow modelling is intricately linked to a broader context. Techniques originally developed for alternative applications are being progressively tailored for passenger flow analysis. As previously demonstrated, this adaptation is occurring at an accelerating pace, particularly with methodologies derived from time series forecasting and machine learning. The full rationale behind this trend remains unexplained and is beyond the scope of this discourse. Assuming the persistence of this trend and its association with time-series modelling and machine learning, it would be insightful to examine which subjects within these spheres are currently prominent.

A trend not yet discernible in quantitative analyses, but which has become prominent in both daily life and academic literature, is the increased utilization of ChatGPT 3.5, grounded in GPT-3.0, as of late 2022 \cite{Brown.2020, Ray.2023}. This discussion transcends the ChatGPT application or the specific GPT architecture, instead focusing on the feasibility of pre-training a model on an extensive dataset and subsequently fine-tuning it for various downstream tasks \cite{Ray.2023}. Although technical capability and the concept of transfer learning, which is the transfer of knowledge from one task to another, are not unprecedented, the performance of such models is fundamentally dependent on scale \cite{Bommasani.2022}. This combination of concepts has culminated in the advent of foundation models, models initially trained on vast datasets that can be fine-tuned for a multitude of downstream applications \cite{Bommasani.2022}. As methodologies from other modelling domains are increasingly adopted for the analysis of short-term passenger flow with progressively reduced time intervals, it is plausible that these foundation models will exert a significant impact. Consequently, models such as Chronos \cite{Ansari.2024}, Moirai \cite{Woo.2024}, Lag-LLama \cite{Rasul.2024}, or TimesFM \cite{Das.2024} may also be applicable.

Although these approaches offer benefits such as cost reduction, enhanced efficiency, and the adaptation of less labeled data, they may also result in the homogenization of models. There exists a potential risk that the unique characteristics and expert insights pertinent to the modelling of short-term passenger flow could be overlooked or inadequately integrated. For instance, amongst the exemplary FMs highlighted previously, only Moirai possesses the capability to directly process multivariate time series, which could be advantageous for synthesizing passenger flow data with auxiliary data. Consequently, the importance of recognizing interdisciplinarity and integrating it into communication processes is increasingly recognised. Beyond modifying existing FMs, it might also be advantageous to develop FMs dedicated to traffic forecasting and to establish specific downstream tasks related to traffic and transit.

The perspectives highlighted above underscore the significant emphasis placed on the enhanced optimization of a model's evaluation criteria. Nevertheless, a model deemed 'perfect' holds limited utility in the absence of tangible benefits. Consequently, it is imperative not to overlook challenges such as efficiency, the constraints in developing entirely customised models ubiquitously, and issues related to data availability. In addition, it is crucial for case studies to be disseminated more extensively on a global scale to facilitate the evaluation of specific study parameters and, where necessary, enable comparative analysis of models in different urban environments.

\section{Conclusion}
\noindent This paper presents the first comprehensive bibliometric analysis of short-term passenger flow forecasting within local public transit, covering 814 publications that span from 1984 to 2024. Our findings indicate that research activity exhibited sporadic patterns prior to 2008, followed by a marked acceleration, characterised by a significant shift from conventional statistical and machine learning methodologies (such as ARIMA, SVM, basic neural networks) towards specialised deep learning architectures (such as LSTM, CNN, and GCN). Rail-based modes predominate in the literature, especially in China, where more than 70\% of the research originates, while studies focused on bus systems are relatively sparse. The dispersion of publication outlets reflects the interdisciplinary character of the domain.  Cocitation networks illustrate an accelerating cross-pollination with broader time-series and machine-learning domains, suggesting that advancements in foundational models (e.g. Chronos, Moirai, Lag-LLama) are poised to influence passenger-flow prediction. However, our review also identifies enduring gaps: constrained data fusion initiatives and underappreciated challenges related to model interpretability, cost-efficiency, privacy, and deployment in real-world settings.

Addressing our three guiding questions, the evolution of topics and structural paradigms reveals a transition from univariate linear forecasting to multivariate, nonlinear, and ultimately deep-learning-based architectures. Prominent contributors and institutions, notably Beijing Jiaotong University and other Chinese universities, have been instrumental in this transformation, while publishing outlets remain highly fragmented across conferences and journals. The implications for future methodologies suggest an adoption of foundational model approaches, indicating that large-scale pre-training and fine-tuning will become increasingly prevalent. Future research should adapt these models to transit-specific downstream tasks while preventing indiscriminate homogenization that overlooks domain expertise.

Our synthesis validates and quantifies insights derived from previous qualitative reviews—particularly the shift towards deep learning—while identifying biases (e.g., geographic, language, mode coverage) and methodological constraints. In conclusion, this study illustrates the field's sustained growth and heightened methodological sophistication, yet it also emphasises the necessity for more open (multivariant) data and an equilibrium between algorithmic performance and practical deployment considerations.

\section{Acknowledgments}
\noindent The project is supported by the Federal Ministry of Transport and Digital Infrastructure (BMVI), grant number 19OI22008A.

\bibliographystyle{IEEEtran}
\bibliography{main.bib}

\appendix
% \ziyue{we also need to add some text descriptions about those figures and tables, what is the insights, etc}

\begin{figure}[H]
\centering
\resizebox{\columnwidth}{!}{\input{fig/co_word_ab_viv.tex}}
\caption{Top 40 abstract trigrams grouped by common attributes. In the mode group modes related to rails dominate with minor relevance of buses as a mode of interest. In comparison, the model group is more diverse and ranges from more traditional time series approaches to more modern deep learning methods. In terms of the type of passenger flows predicted, a distinction is made in some cases, but without a dominant type.}
\label{fig:trigram}
\end{figure}

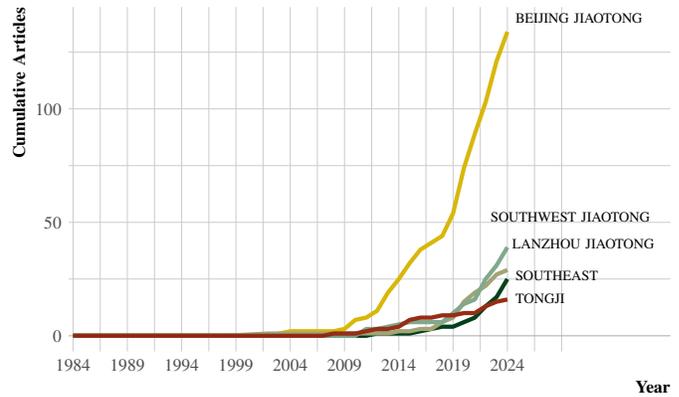
\begin{figure}[H]
\centering
\resizebox{\columnwidth}{!}{% Created by tikzDevice version 0.12.6 on 2025-07-21 12:39:23
% !TEX encoding = UTF-8 Unicode
\begin{tikzpicture}[x=1pt,y=1pt]
\definecolor{fillColor}{RGB}{255,255,255}
\path[use as bounding box,fill=fillColor,fill opacity=0.00] (0,0) rectangle (264.51,180.67);
\begin{scope}
\path[clip] ( 22.44, 18.12) rectangle (264.51,156.76);
\definecolor{drawColor}{gray}{0.80}

\path[draw=drawColor,line width= 0.2pt,line join=round] ( 22.44, 47.26) --
	(264.51, 47.26);

\path[draw=drawColor,line width= 0.2pt,line join=round] ( 22.44, 92.92) --
	(264.51, 92.92);

\path[draw=drawColor,line width= 0.2pt,line join=round] ( 22.44,138.59) --
	(264.51,138.59);

\path[draw=drawColor,line width= 0.2pt,line join=round] ( 35.14, 18.12) --
	( 35.14,156.76);

\path[draw=drawColor,line width= 0.2pt,line join=round] ( 56.98, 18.12) --
	( 56.98,156.76);

\path[draw=drawColor,line width= 0.2pt,line join=round] ( 78.82, 18.12) --
	( 78.82,156.76);

\path[draw=drawColor,line width= 0.2pt,line join=round] (100.66, 18.12) --
	(100.66,156.76);

\path[draw=drawColor,line width= 0.2pt,line join=round] (122.50, 18.12) --
	(122.50,156.76);

\path[draw=drawColor,line width= 0.2pt,line join=round] (144.34, 18.12) --
	(144.34,156.76);

\path[draw=drawColor,line width= 0.2pt,line join=round] (166.18, 18.12) --
	(166.18,156.76);

\path[draw=drawColor,line width= 0.2pt,line join=round] (188.02, 18.12) --
	(188.02,156.76);

\path[draw=drawColor,line width= 0.2pt,line join=round] (209.85, 18.12) --
	(209.85,156.76);

\path[draw=drawColor,line width= 0.2pt,line join=round] (220.77, 18.12) --
	(220.77,156.76);

\path[draw=drawColor,line width= 0.2pt,line join=round] ( 22.44, 24.43) --
	(264.51, 24.43);

\path[draw=drawColor,line width= 0.2pt,line join=round] ( 22.44, 70.09) --
	(264.51, 70.09);

\path[draw=drawColor,line width= 0.2pt,line join=round] ( 22.44,115.76) --
	(264.51,115.76);

\path[draw=drawColor,line width= 0.2pt,line join=round] ( 24.22, 18.12) --
	( 24.22,156.76);

\path[draw=drawColor,line width= 0.2pt,line join=round] ( 46.06, 18.12) --
	( 46.06,156.76);

\path[draw=drawColor,line width= 0.2pt,line join=round] ( 67.90, 18.12) --
	( 67.90,156.76);

\path[draw=drawColor,line width= 0.2pt,line join=round] ( 89.74, 18.12) --
	( 89.74,156.76);

\path[draw=drawColor,line width= 0.2pt,line join=round] (111.58, 18.12) --
	(111.58,156.76);

\path[draw=drawColor,line width= 0.2pt,line join=round] (133.42, 18.12) --
	(133.42,156.76);

\path[draw=drawColor,line width= 0.2pt,line join=round] (155.26, 18.12) --
	(155.26,156.76);

\path[draw=drawColor,line width= 0.2pt,line join=round] (177.10, 18.12) --
	(177.10,156.76);

\path[draw=drawColor,line width= 0.2pt,line join=round] (198.94, 18.12) --
	(198.94,156.76);
\definecolor{drawColor}{RGB}{216,183,10}

\path[draw=drawColor,line width= 1.7pt,line join=round] ( 24.22, 24.43) --
	( 28.59, 24.43) --
	( 63.53, 24.43) --
	( 85.37, 24.43) --
	(102.84, 24.43) --
	(111.58, 26.25) --
	(120.31, 26.25) --
	(124.68, 26.25) --
	(129.05, 26.25) --
	(133.42, 27.17) --
	(137.78, 30.82) --
	(142.15, 31.73) --
	(146.52, 34.47) --
	(150.89, 41.78) --
	(155.26, 47.26) --
	(159.62, 53.65) --
	(163.99, 59.13) --
	(168.36, 61.87) --
	(172.73, 64.61) --
	(177.10, 73.74) --
	(181.46, 92.01) --
	(185.83,105.71) --
	(190.20,118.50) --
	(194.57,134.94) --
	(198.94,146.81);
\definecolor{drawColor}{RGB}{2,64,27}

\path[draw=drawColor,line width= 1.7pt,line join=round] ( 24.22, 24.43) --
	( 28.59, 24.43) --
	( 63.53, 24.43) --
	( 85.37, 24.43) --
	(102.84, 24.43) --
	(111.58, 24.43) --
	(120.31, 24.43) --
	(124.68, 24.43) --
	(129.05, 24.43) --
	(133.42, 24.43) --
	(137.78, 24.43) --
	(142.15, 24.43) --
	(146.52, 25.34) --
	(150.89, 25.34) --
	(155.26, 25.34) --
	(159.62, 25.34) --
	(163.99, 26.25) --
	(168.36, 27.17) --
	(172.73, 28.08) --
	(177.10, 28.08) --
	(181.46, 29.91) --
	(185.83, 31.73) --
	(190.20, 36.30) --
	(194.57, 39.95) --
	(198.94, 47.26);
\definecolor{drawColor}{RGB}{162,164,117}

\path[draw=drawColor,line width= 1.7pt,line join=round] ( 24.22, 24.43) --
	( 28.59, 24.43) --
	( 63.53, 24.43) --
	( 85.37, 24.43) --
	(102.84, 25.34) --
	(111.58, 25.34) --
	(120.31, 25.34) --
	(124.68, 25.34) --
	(129.05, 25.34) --
	(133.42, 25.34) --
	(137.78, 25.34) --
	(142.15, 25.34) --
	(146.52, 25.34) --
	(150.89, 25.34) --
	(155.26, 26.25) --
	(159.62, 26.25) --
	(163.99, 27.17) --
	(168.36, 27.17) --
	(172.73, 29.91) --
	(177.10, 31.73) --
	(181.46, 38.12) --
	(185.83, 41.78) --
	(190.20, 44.52) --
	(194.57, 49.08) --
	(198.94, 50.91);
\definecolor{drawColor}{RGB}{129,168,141}

\path[draw=drawColor,line width= 1.7pt,line join=round] ( 24.22, 24.43) --
	( 28.59, 24.43) --
	( 63.53, 24.43) --
	( 85.37, 24.43) --
	(102.84, 24.43) --
	(111.58, 24.43) --
	(120.31, 24.43) --
	(124.68, 24.43) --
	(129.05, 24.43) --
	(133.42, 24.43) --
	(137.78, 24.43) --
	(142.15, 27.17) --
	(146.52, 27.17) --
	(150.89, 28.08) --
	(155.26, 28.99) --
	(159.62, 29.91) --
	(163.99, 29.91) --
	(168.36, 29.91) --
	(172.73, 29.91) --
	(177.10, 33.56) --
	(181.46, 37.21) --
	(185.83, 39.04) --
	(190.20, 47.26) --
	(194.57, 52.74) --
	(198.94, 60.04);
\definecolor{drawColor}{RGB}{151,45,21}

\path[draw=drawColor,line width= 1.7pt,line join=round] ( 24.22, 24.43) --
	( 28.59, 24.43) --
	( 63.53, 24.43) --
	( 85.37, 24.43) --
	(102.84, 24.43) --
	(111.58, 24.43) --
	(120.31, 24.43) --
	(124.68, 24.43) --
	(129.05, 25.34) --
	(133.42, 25.34) --
	(137.78, 25.34) --
	(142.15, 26.25) --
	(146.52, 27.17) --
	(150.89, 27.17) --
	(155.26, 28.08) --
	(159.62, 30.82) --
	(163.99, 31.73) --
	(168.36, 31.73) --
	(172.73, 32.65) --
	(177.10, 32.65) --
	(181.46, 33.56) --
	(185.83, 33.56) --
	(190.20, 36.30) --
	(194.57, 38.12) --
	(198.94, 39.04);
\definecolor{drawColor}{RGB}{0,0,0}

\path[draw=drawColor,draw opacity=0.10,line width= 0.6pt,line join=round] ( 22.44, 24.43) -- (264.51, 24.43);
\definecolor{drawColor}{RGB}{0,0,0}

\node[text=drawColor,anchor=base west,inner sep=0pt, outer sep=0pt, scale=  0.57] at (202.21, 46.67) {SOUTHEAST};

\node[text=drawColor,anchor=base west,inner sep=0pt, outer sep=0pt, scale=  0.57] at (202.21,150.44) {BEIJING JIAOTONG};

\node[text=drawColor,anchor=base west,inner sep=0pt, outer sep=0pt, scale=  0.57] at (202.21, 37.66) {TONGJI};

\node[text=drawColor,anchor=base west,inner sep=0pt, outer sep=0pt, scale=  0.57] at (192.21, 70.47) {SOUTHWEST JIAOTONG};

\node[text=drawColor,anchor=base west,inner sep=0pt, outer sep=0pt, scale=  0.57] at (200.90, 59.57) {LANZHOU JIAOTONG};
\end{scope}
\begin{scope}
\path[clip] (  0.00,  0.00) rectangle (264.51,180.67);
\definecolor{drawColor}{gray}{0.30}

\node[text=drawColor,anchor=base east,inner sep=0pt, outer sep=0pt, scale=  0.70] at ( 19.56, 22.01) {0};

\node[text=drawColor,anchor=base east,inner sep=0pt, outer sep=0pt, scale=  0.70] at ( 19.56, 67.68) {50};

\node[text=drawColor,anchor=base east,inner sep=0pt, outer sep=0pt, scale=  0.70] at ( 19.56,113.35) {100};
\end{scope}
\begin{scope}
\path[clip] (  0.00,  0.00) rectangle (264.51,180.67);
\definecolor{drawColor}{gray}{0.30}

\node[text=drawColor,anchor=base,inner sep=0pt, outer sep=0pt, scale=  0.70] at ( 24.22, 10.43) {1984};

\node[text=drawColor,anchor=base,inner sep=0pt, outer sep=0pt, scale=  0.70] at ( 46.06, 10.43) {1989};

\node[text=drawColor,anchor=base,inner sep=0pt, outer sep=0pt, scale=  0.70] at ( 67.90, 10.43) {1994};

\node[text=drawColor,anchor=base,inner sep=0pt, outer sep=0pt, scale=  0.70] at ( 89.74, 10.43) {1999};

\node[text=drawColor,anchor=base,inner sep=0pt, outer sep=0pt, scale=  0.70] at (111.58, 10.43) {2004};

\node[text=drawColor,anchor=base,inner sep=0pt, outer sep=0pt, scale=  0.70] at (133.42, 10.43) {2009};

\node[text=drawColor,anchor=base,inner sep=0pt, outer sep=0pt, scale=  0.70] at (155.26, 10.43) {2014};

\node[text=drawColor,anchor=base,inner sep=0pt, outer sep=0pt, scale=  0.70] at (177.10, 10.43) {2019};

\node[text=drawColor,anchor=base,inner sep=0pt, outer sep=0pt, scale=  0.70] at (198.94, 10.43) {2024};
\end{scope}
\begin{scope}
\path[clip] (  0.00,  0.00) rectangle (264.51,180.67);
\definecolor{drawColor}{RGB}{0,0,0}

\node[text=drawColor,anchor=base east,inner sep=0pt, outer sep=0pt, scale=  0.70] at (264.51,  1.36) {\bfseries Year};
\end{scope}
\begin{scope}
\path[clip] (  0.00,  0.00) rectangle (264.51,180.67);
\definecolor{drawColor}{RGB}{0,0,0}

\node[text=drawColor,rotate= 90.00,anchor=base east,inner sep=0pt, outer sep=0pt, scale=  0.70] at (  4.83,156.76) {\bfseries Cumulative Articles};
\end{scope}
\begin{scope}
\path[clip] (  0.00,  0.00) rectangle (264.51,180.67);
\definecolor{drawColor}{RGB}{0,0,0}

\node[text=drawColor,anchor=base west,inner sep=0pt, outer sep=0pt, scale=  0.80] at ( 22.44,163.32) {Based on Corresponding Author $|$ Combined WoS and Scopus data};
\end{scope}
\begin{scope}
\path[clip] (  0.00,  0.00) rectangle (264.51,180.67);
\definecolor{drawColor}{RGB}{0,0,0}

\node[text=drawColor,anchor=base west,inner sep=0pt, outer sep=0pt, scale=  1.00] at ( 22.44,173.77) {\bfseries Affiliations' Production over Time};
\end{scope}
\end{tikzpicture}}
\caption{Cumulative amount of articles grouped by the corresponding author institution. From the top five institutions the Beijing Jiaotong University protrudes compared to the four following also Chinese universities.}
\label{fig:pub_inst}
\end{figure}

\begin{table}[H]
    \caption{Publication channels for documents with the highest cited Total Citation (TC) based on WoS and Scopus data. In order to put the absolute values into context, TC divided by the number of years since publication and normalised total citations (NTC) are also provided. Based purely on the documents with the highest number of citations, TRANSP RES PT C-EMERG TECHNOL stands out as a relevant publication channel.}
    \label{tab:pub_channel}
    \centering
    \resizebox{\columnwidth}{!}{%
    \begin{tabular}{l l c c c c}
        \hline
        Publication Channel & DOI & Year & TC & TC per Year & NTC \\
        \hline
        TRANSP RES PT C-EMERG TECHNOL & 10.1016/j.trc.2011.06.009 & 2012 & 374 & 26.7 & 8.46 \\
        TRANSP RES PT C-EMERG TECHNOL & 10.1016/j.trc.2019.01.027 & 2019 &  281 & 40.1 & 9.89 \\
        NEUROCOMPUTING & 10.1016/j.neucom.2015.03.085 & 2015 &  267 & 24.3 & 10.54 \\
        TRANSP RES PT C-EMERG TECHNOL & 10.1016/j.trc.2017.02.005 & 2017 &  160 & 17.8 & 4.82 \\
        TRANSP RES PT C-EMERG TECHNOL & 10.1016/j.trc.2019.08.005 & 2019 &  158 & 22.6 & 5.56 \\
        \hline
    \end{tabular}%
    }
\end{table}
\begin{table}[H]
    \caption{Missing data from different tags of the two queries. Scopus has a total of 784, WoS 784 result, and combined 821 results. In context of the total amount of data points every category until corresponding author can be assumed to have a small fraction of missing data. 11.08\% of the corresponding authors and 19.36\% of the keywords are missing resulting in a bias for analysis using those entries. Keywords plus, cited references and science categories are not used in this bibliometric analysis and are just listed for sake of completeness. }
    \label{tab:missing_data}
    \centering
    \begin{tabular}{lrrr}
        \hline
        Description & Combined & Scopus & WoS\\
        \hline
        Author                    & 0   & 1 & 0  \\
        Document Type             & 0   & 0 & 0 \\
        Language                  & 0   & 0 & 0 \\
        Publication Year          & 0   & 0 & 0 \\
        Title                     & 0   & 0 & 0 \\
        Total Citation            & 0   & 0 & 0 \\
        Journal                   & 1   & 1 & 0 \\
        Affiliation               & 5   & 4 & 5  \\
        Abstract                  & 10  & 2 & 10 \\
        Corresponding Author      & 91  & 164 & 5 \\
        DOI                       & 93  & 34 & 72 \\
        Keywords                  & 159 & 158 & 107 \\
        Keywords Plus             & 193 & 89 & 164 \\
        Cited References          & 821 & 784 & 2 \\
        Science Categories        & 821 & 784 & 508 \\
        \hline
    \end{tabular}
\end{table}
\newpage

\begin{IEEEbiography}[{\includegraphics[width=1in,height=1.25in,clip,keepaspectratio]{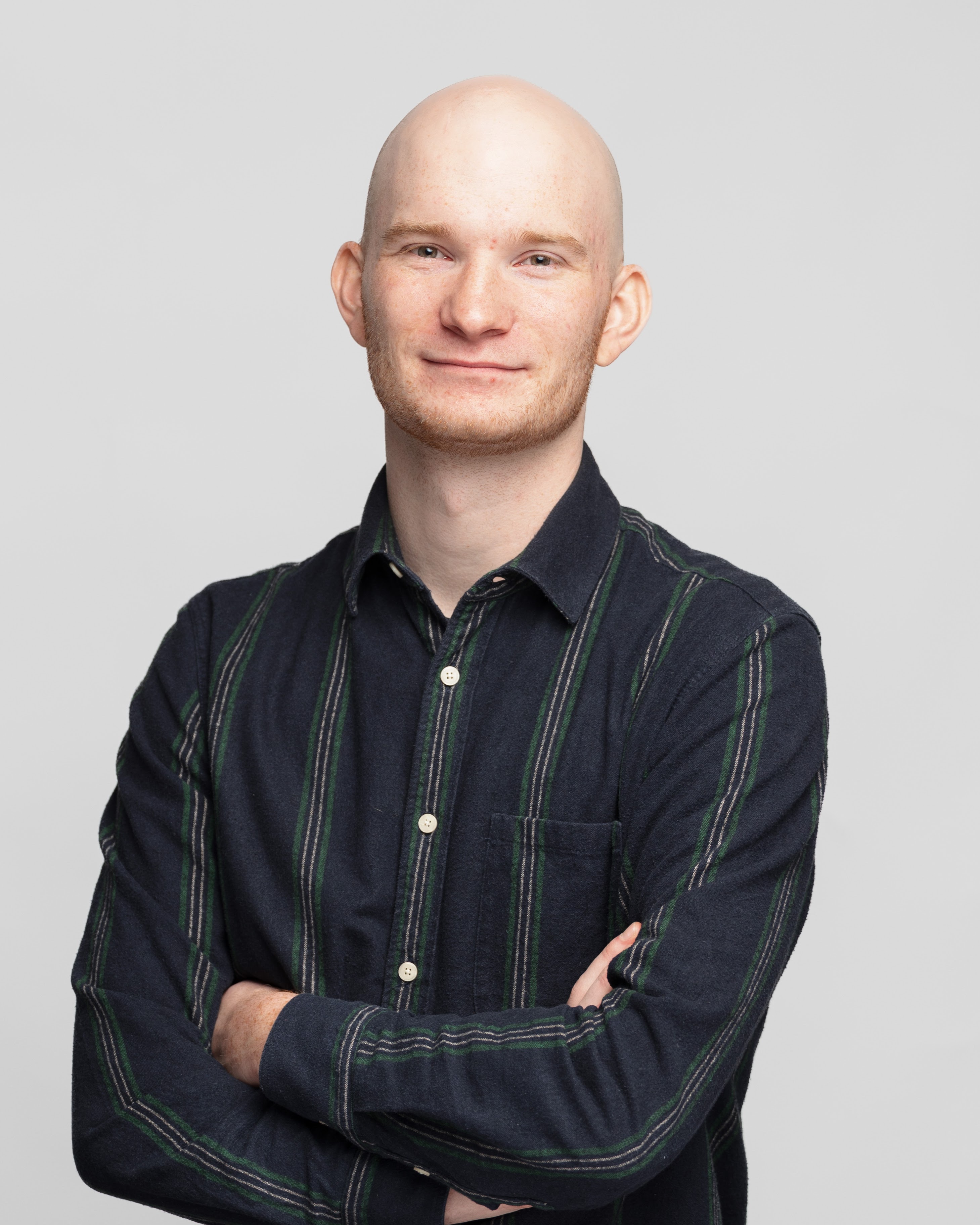}}]{Jonathan Hecht} is a master's student at HafenCity University Hamburg and a research assistant in the research group of Prof. Youness Dehbi. His research focuses on data-driven approaches to analysing urban mobility patterns, with a particular interest in geospatial data integration and machine learning for mobility applications. He has professional experience in the mobility sector, particularly in active mobility and sustainable transportation systems.
\end{IEEEbiography}

\begin{IEEEbiography}[{\includegraphics[width=1in,height=1.25in,clip,keepaspectratio]{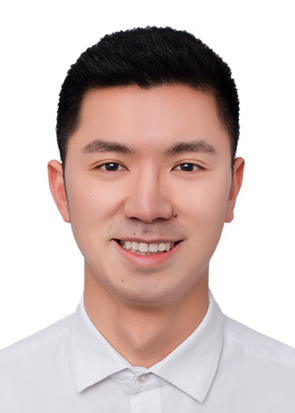}}]{Weilian Li}
is an Associate Professor at the Faculty of Geosciences and Engineering at Southwest Jiaotong University, Chengdu, China. His research primarily focuses on the application of Geographic Information Systems (GIS) in disaster management, but also methodological expertise in Machine Learning. He has a strong interest in topics related to smart cities, disaster management, and digital twins. He serves as the secretary of the ISPRS ICWG III/IVa on Disaster Management and representative of the UN-GGIM Academic Network.
\end{IEEEbiography}

\begin{IEEEbiography}[{\includegraphics[width=1in,height=1.25in,clip,keepaspectratio]{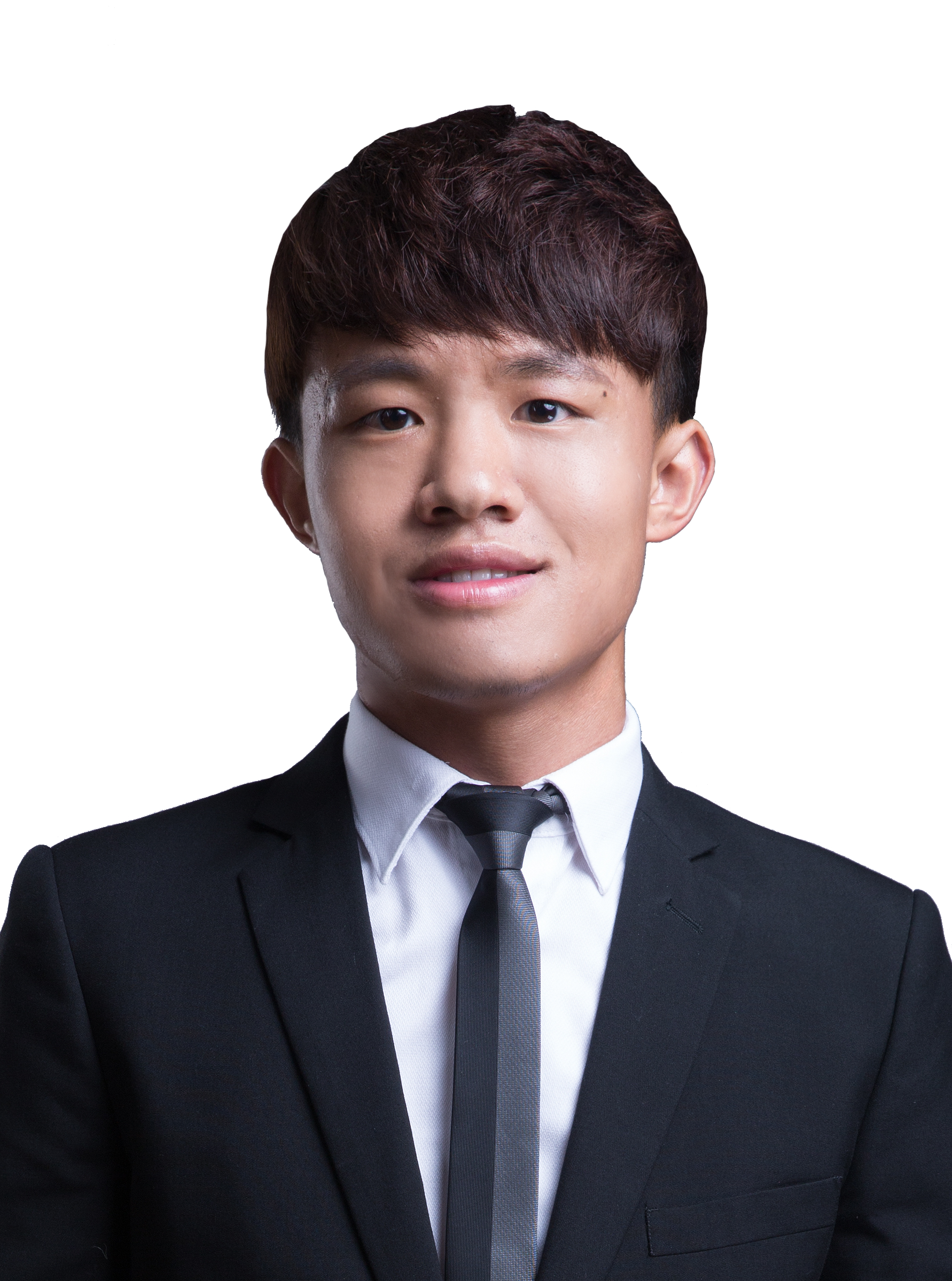}}]{Ziyue Li}
is a W2 professor at the Heilbronn Data Science Center and Department of Operations \& Technology, Technical University of Munich. His research targets high-dimensional data mining and deep learning methodologies for real-world spatiotemporal problems. His expertises are tensor analysis, spatiotemporal data, and statistical machine learning. Those methods have been applied to various industries, mainly in smart transportation. Dr. Li's works have been awarded various Best Paper awards in INFORMS, IISE, and IEEE CASE.
\end{IEEEbiography}

\begin{IEEEbiography}[{\includegraphics[width=1in,height=1.25in,clip,keepaspectratio]{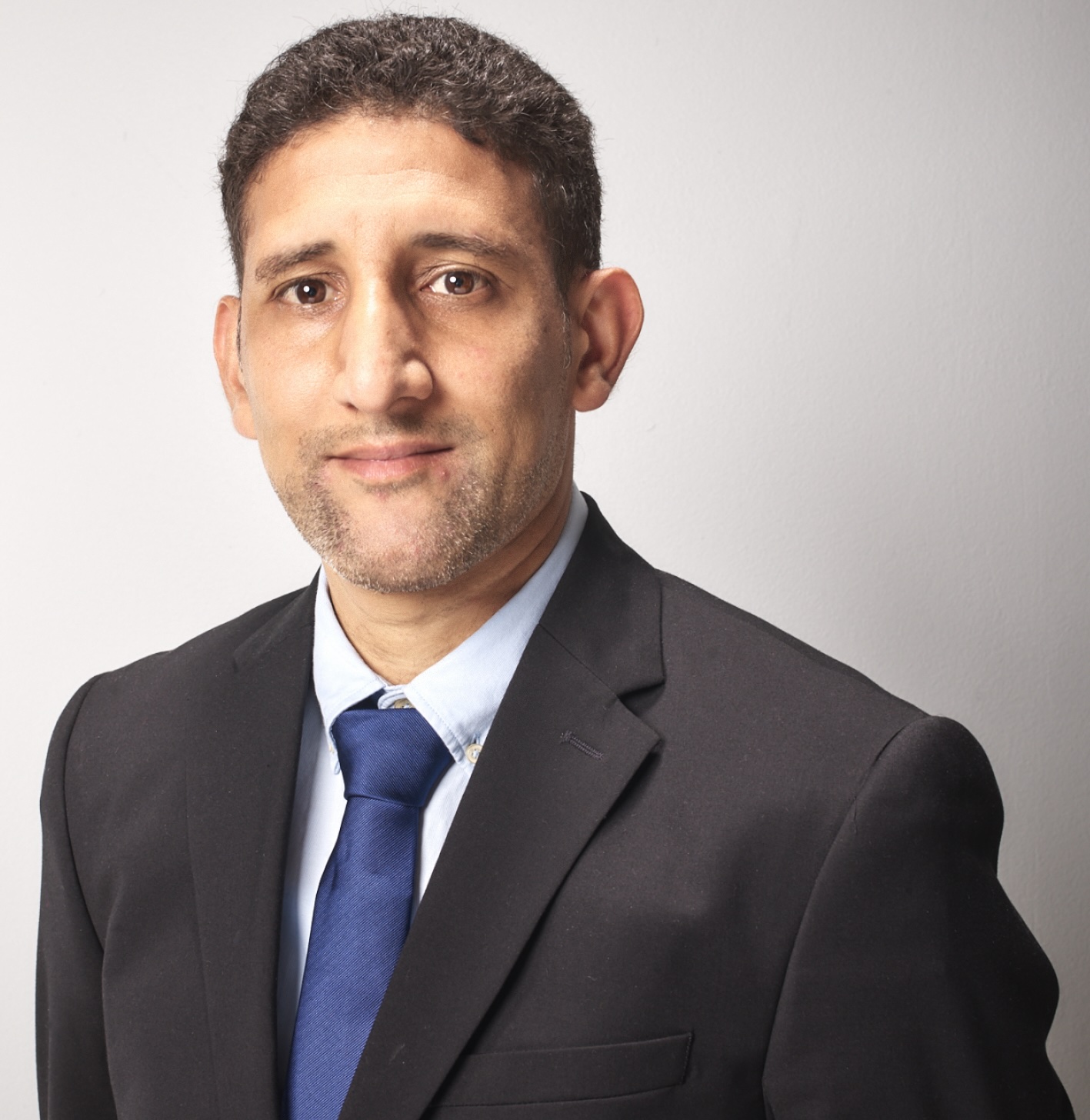}}]{Youness Dehbi} 
is a Full Professor of Computational Methods at HafenCity University Hamburg. His research focuses on developing and applying methods and algorithms for the analysis of spatial and temporal problems. His work centers on digital twins, smart cities, Building Information Modeling (BIM), and Geographic Information Systems (GIS), emphasizing their integration. A key goal of his research is to intelligently link geospatial data with urban infrastructures to drive digital transformation in planning, construction, and mobility. %Dr. Dehbi has received multiple awards for his scientific contributions, including Best Paper Awards at international conferences such as Smart Data and Smart Cities and 3DGeoInfo.
\end{IEEEbiography}

% \newpage
% \begin{figure*}[tbp]
% \centering
% \resizebox{\textwidth}{!}{\input{fig/topic_cluster.tex}}
% \caption{Clustered co-occurrence patterns via a Biterm Topic Model (BTM). Each clusters is related to some content topic. Higher percentage depict a higher probability for such a word combination in the corpus of documents.Cluster 7 is removed due to a low percentage value.}
% \label{fig:btm}
% \end{figure*}

\vfill

\end{document}